\documentclass[fleqn,10pt]{SelfArx} 

\usepackage[english]{babel} 

\usepackage{lipsum} 
\usepackage{afterpage}

 \definecolor{color1}{RGB}{0,0,110} 
\definecolor{color2}{RGB}{0,20,80} 

\usepackage{hyperref} 
\hypersetup{hidelinks,colorlinks,breaklinks=true,urlcolor=color2,citecolor=color1,linkcolor=color1,bookmarksopen=false,pdftitle={Title},pdfauthor={Author}}

\JournalInfo{Uploaded to arXiv.org} 
\Archive{} 

\PaperTitle{Scaling Behavior of Bipolar Nanopore Rectification with Multivalent Ions} 

\Authors{D\'avid Fertig\textsuperscript{1},Bart\l{}omiej Matejczyk\textsuperscript{2}, M\'onika Valisk\'o\textsuperscript{1}, Dirk Gillespie\textsuperscript{3},  Dezs\H{o} Boda\textsuperscript{1}*} 
\affiliation{\textsuperscript{1}\textit{Department of Physical Chemistry, University of Pannonia,  P. O. Box 158, H-8201 Veszpr\'em, Hungary}} 
\affiliation{\textsuperscript{2}\textit{Department of Mathematics, University of Warwick, CV4 7AL Coventry, United Kingdom}}%
\affiliation{\textsuperscript{3}\textit{Department of Physiology and Biophysics, Rush University Medical Center, Chicago, Illinois 60612-3833, USA}}
\affiliation{*\textbf{Corresponding author}: dezsoboda@gmail.com} 

\Keywords{nanopore --- scaling --- Nernst-Planck --- Monte Carlo} 
\newcommand{\keywordname}{Keywords} 

\Abstract{We present a scaling behavior of a rectifying bipolar nanopore as a function of the parameter $\xi=R_{\mathrm{P}}/(\lambda z_{\mathrm{if}})$, where $R_{\mathrm{P}}$ is the radius of the pore, $\lambda$ is the characteristic screening length of the electrolyte filling the pore, and $z_{\mathrm{if}}=\sqrt{z_{+}|z_{-}|}$ is a scaling factor that makes scaling work for electrolytes containing multivalent ions ($z_{+}$ and $z_{-}$ are cation and the anion valences).
By scaling we mean that the rectification of the pore (defined as the ratio of currents in the forward and reversed biased states) depends on pore radius, concentration, $c$, and ion valences via the parameter $\xi$ implicitly.
This feature is based on the fact that rectification depends on the voltage-sensitive appearance of depletion zones that, in turn, depend on the relation of $R_{\mathrm{P}}$ to the rescaled screening length $\lambda z_{\mathrm{if}}$.
In this modeling study, we use the Poisson-Nernst-Planck (PNP) theory and a particle simulation method, the Local Equilibrium Monte Carlo (LEMC).
The latter can compute ion correlations that are ignored in the mean-field treatment of PNP and that are very important for multivalent ions (we show results for 1:1, 2:1, 3:1, and 2:2 electrolytes).
In addition to the $z_{\mathrm{if}}$ factor, we show that one must choose a screening length appropriate to the system, in our case the Debye length for $\lambda$ for PNP and the screening length given by the Mean Spherical Approximation for LEMC.}

\begin{document}

\flushbottom 

\maketitle 


\thispagestyle{empty} 


\section{Introduction}
\label{sec:intro}

Nanopores are often defined superficially as pores whose radius, $R_{\mathrm{P}}$, falls into the nanometer range.
A more functional definition, however, is that nanopores are distinguished from micropores by the feature that their radius is comparable to the characteristic screening length, $\lambda$, of the electrolyte with which we fill the pore.
The screening length, $\lambda$, (to be specified later) depends on the concentration of the electrolyte, $c$, properties of ions (e.g., valences, $z_{i}$, and radii, $R_{i}$), and properties of the solvent (e.g., dielectric constant, $\epsilon$), namely
\begin{equation}
 \lambda = \lambda(c,z_{+},z_{-},R_{+},R_{-},\epsilon)
\end{equation} 
for a given temperature.
This definition binds the geometrical features of the pore to the properties of the electrolyte, the medium in which charge transport takes place.

In this paper, we quantify this statement and show that the scaling parameter of the ratio of pore radius to screening length, $R_{\mathrm{P}}/\lambda$, is a factor that determines some pore behaviors. 
Scaling means that the device function depends only on a combined parameter (e.g., $R_{\mathrm{P}}/\lambda$) that is put together from other variables (e.g., $R_{\mathrm{P}}$, $c$, $z_{+}$, $z_{-}$); that is, it depends on these only implicitly via the combined parameter.

The scaling $R_{\mathrm{P}}/\lambda$ is not new in principle or in practice. 
Many studies have used the ratio of pore radius to screening length to describe various aspects of fluidic pore behavior\cite{daiguji_nl_2005,vlassiouk_nl_2007,yan_nl_2009,albrecht_chapter_2013,Abgrall_2008,bocquet_csr_2010,daiguji_csr_2010,eijkel_csr_2010,zangle_csr_2010,dal_cengio_jcp_2019}.
Here, we introduce two generalizations that extend the scaling idea to electrolytes with multivalent ions. 
First, we pick a screening length that is appropriate to the system being considered. 
Specifically, we generally do not use the Debye length, $\lambda_{\mathrm{D}}$, like previous studies. 
Electrolytes, especially those with multivalent ions, are not well described by the Poisson-Boltzmann (PB) theory from which the Debye length is derived. 
Therefore, it is generally not the best choice of screening length. 

Second, we include an additional factor of $z_{\mathrm{if}}=\sqrt{z_{+}|z_{-}|}$, previously derived from field-theoretic arguments \cite{dicaprio_mp_2006}, to modify the screening length. 
As we will show in Section \ref{sec:multivalent}, the appropriate scaling parameter that is valid even for multivalent electrolytes is
\begin{equation}
 \xi =\dfrac{R_{\mathrm{P}}}{\lambda z_{\mathrm{if}}}.
 \label{eq:xi}
\end{equation} 
Here, we show that rectification in a bipolar nanopore can be described by the scaling parameter $\xi$. 
Rectification is the ratio of the currents in the ON and OFF (forward and reverse biased) states of the nanopore:
\begin{equation}
 r=\dfrac{I^{\mathrm{ON}}}{I^{\mathrm{OFF}}} ,
 \label{eq:totrect}
\end{equation} 
where $I^{\mathrm{ON}}=I(200 \;\mathrm{mV})$ and $I^{\mathrm{OFF}}=|I(-200\; \mathrm{mV})|$ are the absolute values of the total currents in the ON and OFF states, respectively, where $200$ mV is a representative value of the voltage.
(When talking about current values in this work, we always mean absolute values, even though currents are negative for negative voltages.)

Scaling means that a smooth function 
\begin{equation}
 r=r(\xi)
\end{equation}
exists and that $r$ is the same for different combinations of $R_{\mathrm{P}}$, $c$, $R_{+}$, $R_{-}$, $z_{+}$, and $z_{-}$ when $\xi$ is the same for these combinations.
We show this first for 1:1 electrolytes and then extend it to 2:1, 3:1, and 2:2 electrolytes.
We present scaling behavior in the parameter space of pore radius, $R_{\mathrm{P}}$, concentration, $c$, and ion valences, $z_{+}$ and $z_{-}$ (cations and anions will be denoted by $+$ and $-$, respectively).

Such scaling behavior is both a way to understand physics of how a device operates and a practical tool.
Imagine that we have a nanopore of radius $R_{\mathrm{P}}$ and electrolytes of concentrations $c$.
Measuring the device function ($r$, for example) for a series of concentrations, we can establish the function $r(\xi)$. 
This makes it possible to predict $r$ for another pore radius $R'_{\mathrm{P}}$, an unstudied electrolyte concentration $c'$, or a completely different electrolyte, all without actually fabricating the nanopore of radius $R'_{\mathrm{P}}$ or mixing new electrolytes.
Because fabrication and experiments are expensive and/or difficult, the predictive power of such a simple formula can be very useful in the design of nanodevices.

In this study, we show our proposed scaling for rectification by simulating the nanopore and its ionic current.
The electrolyte is modeled in the implicit solvent framework, an approximation common in nanopore modeling studies \cite{daiguji_nl_2005,constantin_pre_2007,wolfram_jpcm_2010,Ali_ACSnano_2012,gamble_jpcc_2014,nikolaev_jce_2014} and one that captures the device-level physics \cite{hato_pccp_2017,valisko_jcp_2019}.

This work was inspired by our previous study in which we found the presence of scaling \cite{madai_pccp_2018}.
A symmetric charge pattern was used in that study for a model nanofluidic transistor, where current was modulated with the surface charge of the central region.
Defining the ON and OFF states of the transistor at characteristic values of that surface charge, we defined the device function as the ratio of the ON and OFF currents quantifying switching.
We showed that the device function scaled with $R_{\mathrm{P}}/\lambda_{\mathrm{D}}$ for a 1:1 electrolye.
In that study \cite{madai_pccp_2018}, we used the Debye length to characterize screening in the electrolyte:
\begin{equation}
\lambda_{\mathrm{D}} = \left( \sum_{i} \dfrac{e^{2}z_{i}^{2}c_{i}}{\epsilon_{0}\epsilon kT} \right)^{-1/2},
\label{eq:lambdaD}
\end{equation} 
where $e$ is the unit charge, $c_{i}$ is the bulk concentration of ionic species $i$, $\epsilon_{0}$ is the permittivity of vacuum, $k$ is Boltzmann's constant, and $T$ is the absolute temperature.
Basically, for a given electrolyte system ($z_{+}{:}z_{-}$), the Debye length increases with decreasing concentration.
In the simple electrolyte considered here, ion concantrations are related to salt concentration via $c=c_{+}/|z_{-}|=c_{-}/z_{+}$.

The relation of the pore dimension and the screening length was discussed in several experimental and modeling works \cite{daiguji_nl_2005,vlassiouk_nl_2007,yan_nl_2009,albrecht_chapter_2013,Abgrall_2008,bocquet_csr_2010,daiguji_csr_2010,eijkel_csr_2010,zangle_csr_2010,dal_cengio_jcp_2019}.
Note that there are length scales used to characterize nanofluidic devices beyond pore radius and electrolyte screening length discussed, for example, by Bocquet and Charlaix \cite{bocquet_csr_2010}.
In our paper we do not change the length of the pore, for example; we leave that to later studies.
The surface charge is also fixed in this work; the length scale associated with surface charge is the Dukhin length characterizing pore width below which surface conduction dominates over the bulk conductance.

This shows that the idea of finding simple relations between basic characteristics of the nanopore is quite old.
The merit of this study is that we provide a quantitative analysis of scaling with the new $\xi$ parameter which also encompasses multivalent ions.

The importance of multivalent ions in achieving peculiar conductance behavior of nanopores due, for example, to charge inversion or charge selectivity, is well known \cite{he_jacs_2009,garciagimenez_pre_2010,gurnev_langmuir_2012,li_nl_2015,ramirez_jmembsci_2018,mashayak_jcp_2018,chou_nl_2018,voukadinova_jcp_2019}. 
Still, the number of papers dealing with multivalent electrolytes (either $z_{+}>1$ or $|z_{-}|>1$)   compared to NaCl or KCl is relatively small, also see references \cite{kuo_langmuir_2001,ho_pnas_2005,albesa_jmm_2013,rollings_natcomms_2016,rangharajan_micronano_2016,nandigana_scirep_2018,wang_jphyschemc_2018,nasir_jcis_2019,alidoosti_electrophoresis_2019,liu_nnl_2019,li_jphyschemc_2019,zqli_jphyschemc_2019,zhang_analchem_2019}.

Multivalent electrolytes are also interesting from a modeling point of view.
Strong ionic correlations appear in these systems, but are poorly accounted for by the mean-field PB theory and its non-equilibrium counterpart, the Poisson-Nernst-Planck (PNP) theory.
This theory uses the Nernst-Planck (NP) equation to describe ion transport:
\begin{equation}
 \mathbf{j}_{i}(\mathbf{r})=-\dfrac{1}{kT}D_{i}(\mathbf{r})c_{i}(\mathbf{r}) \nabla \mu_{i}(\mathbf{r}) ,
\label{eq:np}
\end{equation} 
where $\mathbf{j}_{i}(\mathbf{r})$ is the particle flux density of ion species $i$, $c_{i}(\mathbf{r})$ is the concentration profile, $\mu_{i}(\mathbf{r})$ is the electrochemical potential profile, and $D_{i}(\mathbf{r})$ is the diffusion coefficient profile. 
Here, we use both PB theory and a particle simulation technique, Local Equilibrium Monte Carlo (LEMC) \cite{boda_jctc_2012}, to compute $\mu_{i}(\mathbf{r})$ and $c_{i}(\mathbf{r})$.

These methodologies are outlined briefly in the next section and described in detail in our earlier papers  \cite{boda_jctc_2012,boda_jml_2014,matejczyk_jcp_2017}. 
The important difference is that LEMC can reproduce ionic correlations, so it is reasonable to apply it for multivalent electrolytes in order to quantify the errors introduced by the mean-field treatment of PNP.
As we will show, using a screening length that also includes these correlations (e.g., from the Mean Spherical Approximation (MSA)) is a better choice in the case of LEMC, while $\lambda_{\mathrm{D}}$, which is a product of the PB theory, is a natural choice in PNP.

While a PNP vs.\ LEMC comparative analysis with special attention to charge inversion will be published in our subsequent work, the deviations between PNP and LEMC will be apparent in this study as well, where we apply both methods to study scaling.

\section{Models and methods}
\label{sec:model}

\begin{figure}[t]
\begin{center}
\includegraphics*[width=0.33\textwidth]{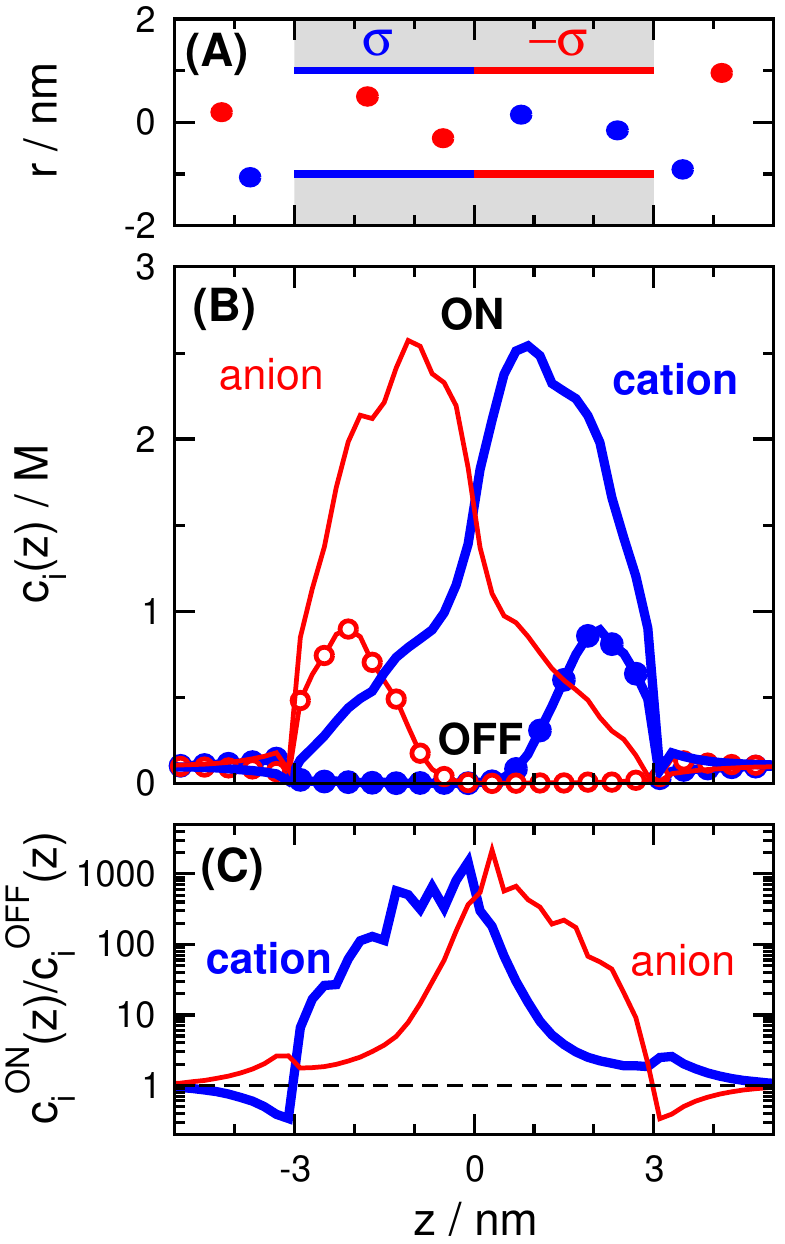}
\caption{
(A) The schematics of the nanopore. The pore's radius is $R_{\mathrm{P}}$ (varying parameter), while its length is $6$ nm (fixed parameter). There is $\sigma=1$ $e$/nm$^{2}$ surface charge on the pore wall on the left hand side ($z<0$), while there is $-\sigma$ surface charge on the pore wall on the right hand side ($z>0$).  Anions and cations are indicated by red and blue circles, respectively.
(B) The concentration profiles of a 1:1 electrolyte for the ON ($200$ mV) and OFF ($-200$ mV) states as obtained from LEMC simulations. Thick blue and thin red lines refer to cations and anions, respectively. The curves with the symbols refer to the OFF state ($R_{\mathrm{P}}=1$ nm, $c=0.1$ M).
(C) The ratio of the ON- and OFF-state concentration profiles.
}\label{fig1}
\end{center}
\end{figure}

We consider a cylindrical bipolar nanopore with negative and positive surface charges ($\pm\sigma=\pm 1$ $e\,$nm$^{-2}=\pm 0.16$ Cm$^{-2}$) on the pore wall in the two half regions along the pore axis ($z$ coordinate) as shown in Fig.\ \ref{fig1}A.
The length of the pore is fixed at $6$ nm, while the radius, $R_{\mathrm{P}}$, changes.
The walls of the pore and the membrane are hard walls forbidding the overlap of hard-sphere ions with them.

Water is treated implicitly in our model which means that its two major effects on ions are modeled by two response functions. 
One effect is an ``energetic'' one: the screening of the Coulomb potential acting between ions.
It is taken into account by the dielectric constant, $\epsilon$, of the continuum dielectric in the denominator of the ion-ion potential:
\begin{equation}
 u_{ij}(r)
=\left\{
        \begin{array}{ll}
    \infty & \; \mbox{for} \; \;  r<R_{i}+R_{j}\\
        \dfrac{z_{i}z_{j}e^{2}}{4\pi \epsilon_{0}\epsilon r} & \; \mbox{for} \; \; r \geq R_{i}+R_{j}  ,
        \end{array}
        \right. 
\label{eq:pm}
\end{equation} 
where $r$ is the distance between the ions.
The hard sphere component is absent in the PNP calculations, as are electrostatic correlations beyond the mean-field.

The other effect is a ``dynamic'' one: the diffusion of ions is hindered by water via friction.
Diffusion is also limited by interactions with other ions and the confining pore.
This is taken into account by the diffusion coefficient profile, $D_{i}(\mathbf{r})$, of ions in the NP equation (Eq.\ \ref{eq:np}).
The diffusion coefficient is a user-specified parameter.
While its value could be extracted from all-atom molecular dynamics (MD) simulations, it is more common to consider it as an adjustable parameter and to fit its value either to MD results \cite{hato_pccp_2017,valisko_jcp_2019} or to experimental data \cite{gillespie-jpcb-109-15598-2005,gillespie_bj_2008_energetics,gillespie_bj_2008,gillespie_bj_2008b,boda_jgp_2009,boda_arcc_2014}.
In this pure theory study, the particular choice does not qualitatively affect our conclusions so long as the pore is the highest resistance element. 
We used an infinite dilution value of $1.334\cdot10^{-9}$ m$^{2}$/s for both ions in the bulk, while the value inside the pore is ten times smaller as in our earlier works \cite{matejczyk_jcp_2017,madai_jcp_2017,madai_pccp_2018}.
The value inside the pore just scales the current and do not influence rectification, because it scales the current in the ON and OFF states the same way.

For the ionic radii we used $R_{+}=R_{-}=0.15$ nm.
This means that cations and anions behave the same way in 1:1 and 2:2 electrolytes in our nanopore model where the left and right halves are identical except the sign of the surface charge (Fig.\ \ref{fig1}A).

The statistical mechanical methods with which we compute the relation between $c_{i}(\mathbf{r})$ and $\mu_{i}(\mathbf{r})$ in the NP equation (Eq.\ \ref{eq:np}) are the LEMC simulation method and PB theory.
Coupled to the NP equation, they form the NP+LEMC and PNP methods, respectively.

The LEMC technique is a grand canonical simulation devised for a non-equilibrium situation.
This means that the chemical potential is not constant system-wide (as it would be in equilibrium), but is a function of position.
We divide the system into small volume elements, assume local equilibrium in them, and apply particle insertion and deletion steps with the same formula for acceptance probabilities as in equilibrium simulations, but using the local chemical potential of the volume element.
An LEMC run provides the $c_{i}(\mathbf{r})$ profile as an output for the $\mu_{i}(\mathbf{r})$ profile, which was the input.
In the coupled NP+LEMC method, the $\mu_{i}(\mathbf{r})$  profile is changed in an iterative way, until the $\mathbf{j}_{i}(\mathbf{r})$ flux density resulting from the $c_{i}(\mathbf{r})$ and $\mu_{i}(\mathbf{r})$ profiles satisfies conservation of mass ($\nabla\cdot\mathbf{j}_{i}(\mathbf{r})=0$).
The LEMC method correctly computes volume exclusion and electrostatic correlations between ions, so it goes beyond the mean-field description of the PNP theory.

In the PNP theory, the mean electrical potential profile, $\Phi(\mathbf{r})$, is computed from the charge profile, $\sum_{i}z_{i}ec_{i}(\mathbf{r})$, by solving Poisson and NP equations (Eq.\ \ref{eq:np}) with the electrochemical potential $\mu_{i}^{\mathrm{PNP}}(\mathbf{r})=\mu_{i}^{0}+kT\ln c_{i}(\mathbf{r})+z_{i}e\Phi(\mathbf{r})$.
The latter means that the electrolyte solution is ideal: the excess part of $\mu_{i}(\mathbf{r})$ contains only the mean-field term, $z_{i}e\Phi(\mathbf{r})$, while extra ionic correlation terms beyond this, including volume exclusion effects, are ignored.

These extra ionic correlation effects beyond mean field are naturally included in LEMC, while ions are point charges in PNP ``feeling'' only the effect of the mean electrical potential.
PNP is a continuum theory, where the distribution of ions is described by continuous functions (density profiles).
LEMC, on the other hand, moves ions explicitly as particles and samples the configurational space, $\{\mathbf{r}^{N}\}$, by considering actual configurations of ions in the three-dimensional simulation cell.
Spatial averages of the outcome produce continuous concentration profiles.

The simulation cell is a cylinder with the membrane at $z=0$ which is much wider and longer than the nanopore.
The system is rotationally symmetric.
Boundary conditions are applied at the two half cylinders on the two sides of the membrane for concentrations and the electrical potential as described in earlier works \cite{boda_jctc_2012,boda_jml_2014, matejczyk_jcp_2017}.
The Dirichlet boundary conditions for the electrical potential model the electrodes.
In our study, the concentration is the same in the left and right baths, while a voltage is applied across the membrane that is the driving force of the steady-state ionic flux.

The surface charges, $\pm\sigma$, on the wall of the nanopore are modeled as fractional point charges on a grid in LEMC, while they are taken into account by Neumann boundary conditions in PNP.

\section{Results}

\subsection{Mechanism of rectification}

Rectification is governed by the voltage-dependent appearance of depletion zones of the coions in the two regions (cations in the positive region and anions in the negative region) as shown in Fig.\ \ref{fig1}B for a 1:1 electrolyte.
The purpose of this figure is to illustrate the mechanism of rectification in a bipolar nanopore. 
Depletion zones are regions along the $z$-axis where the individual ionic concentrations are low in the OFF state compared to the ON state.
If we imagine the nanopore along the $z$-axis as resistors connected in series, a high-resistance element for a given ionic species makes the resistance of the whole nanopore high.
Depletion zones, therefore, determine the conductivity behavior of the pore: deeper depletion  zones mean larger resistance and smaller current.

Figure \ref{fig1}B shows that depletion zones are deeper in the OFF state ($-200$ mV) than in the ON state ($200$ mV).
To emphasize that the depletion zone is a relative concept (OFF vs.\ ON), we plot a relative profile obtained by dividing the ON-state concentration profile, $c_{i}^{\mathrm{ON}}(z)$, with the OFF-state concentration profile, $c_{i}^{\mathrm{OFF}}(z)$, in Fig.\ \ref{fig1}C. 
This relative profile characterizes the presence of such depletion zones in the OFF state relative to the ON state and the degree of rectification exhibited by the pore.
This profile has a large peak at the depletion zones of both ions, which means that current is much larger in the ON state for both ions. Therefore, rectification is present for the total current.

More detailed analyses about the mechanism of rectification and how it is primarily controlled by axial ($z$-dependent) concentration profiles are given in our previous studies \cite{hato_pccp_2017,valisko_jcp_2019,matejczyk_jcp_2017}.
In particular, molecular dynamics (MD) simulations of all-atom models including explicit water were also performed for bipolar nanopores \cite{hato_pccp_2017} and nanopores with varying charge patterns \cite{valisko_jcp_2019}.
In those works, we showed that  the implicit-water model used here reproduces the device behavior given by the explicit-water model because it can reproduce the qualitative behavior of the axial concentration profiles, even if it misses details in the radial direction.
Although, in general, MD simulations would be preferable,  they have limitations at small concentrations and are slower. 
Reduced models with their ability to capture overall device physics (i.e., the physics that governs how inputs like voltage and concentrations become measurable outputs like current) are a good, computationally tractable alternative.

\subsection{The relation of depletion zones and double layer overlap}
\label{sec:overlap}

Depletion zones, therefore, are the key characteristics of nanopores whose behavior is controlled by surface charge patterns and external field.
These nanopores are the fluidic cousins of solid state semiconductor devices, where depletion zones of electrons and holes are tuned by doping and external electric field.
Here, the role of doping is played by the surface charges on the pore wall. 
The surface charge can be fixed, produced by pH-dependent protonated/deprotonated groups, or can be polarization charge, produced by an external electric field on the surface of a metal.

\begin{figure}[t!]
	\begin{center}
		\includegraphics*[width=0.47\textwidth]{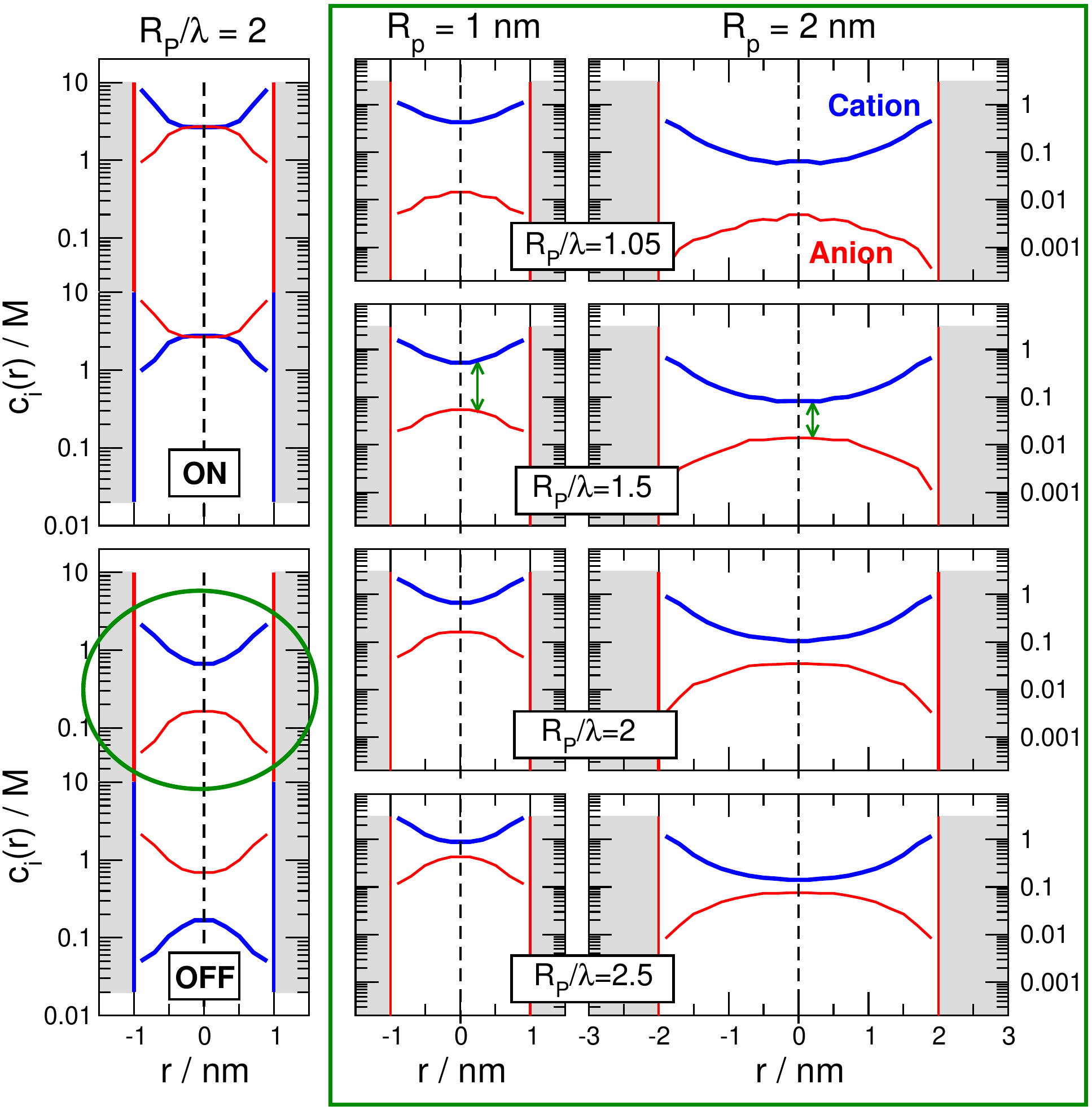}
		\vspace{0.5cm}
		\caption{
		Illustration of the behavior of the electric double layers in nanopores for different values of the parameter $R_{\mathrm{pore}}/\lambda$ via radial concentration profiles. 
		These profiles are computed by the NP+LEMC method, so $\lambda=\lambda_{\mathrm{MSA}}$.
		The profiles are obtained by averaging over the negative (indicated by vertical red lines) and positive (indicated by vertical blue lines) regions in the $z$-direction. 
		Profiles are shown for a 1:1 electrolyte.
		The left column shows the curves for $R_{\mathrm{P}}/\lambda_{\mathrm{MSA}}=2$ for both regions in the ON (top) and the OFF (bottom) states ($R_{\mathrm{P}}=1$ nm). 
		Because the electrolyte is symmetric, the cation and anion profiles are interchangable in the two regions. 
		Since the OFF state is the important state from the point of view of rectification, we plot the case circled by a green ellipse in the middle and right panels (in the large green rectangle) only for the negatively charged region, where the cation is the counterion and the anion is the coion.
		Results are shown for pore radii $R_{\mathrm{P}}=1$ nm (middle column) and $R_{\mathrm{P}}=2$ nm (right column) for $R_{\mathrm{P}}/\lambda_{\mathrm{MSA}}=1.05$, $1.5$, $2$, and $2.5$ (from top to bottom).
 		The concentrations that correspond to these state points can be found in Table S1 of the SI (they are in the $c=0.02968-1.096$ M concentration range).
		Thick blue lines refer to cations, while thin red lines refer to anions. 
				}
		\label{fig2}
	\end{center}
\end{figure}

The key feature behind the scaling behavior is that the depth of these depletion zones is strongly associated with the overlap of the  double layers that are formed at the wall of the nanopore in the radial ($r$) dimension.
This is illustrated in Fig.\ \ref{fig2} for a 1:1 electrolyte.

The left column shows the radial profiles for both zones averaged over the given zone in the $z$-dimension in the ON (top) and OFF (bottom) states for the value $R_{\mathrm{P}}/\lambda=2$  (In this figure we show NP+LEMC results, so we use the $\lambda_{\mathrm{MSA}}$ values for $\lambda$ introduced later in section \ref{sec:msa}. That said, we will use $\lambda$ for a general discussion of Fig.\ \ref{fig2} in the main text.).
For negative (positive) surface charge the cations (anions) are the counterions, while the anions (cations) are the coions.
In the ON state, the concentration in the centerline of the pore ($r=0$) is large, forming a bulk-like fluid with cation and anion concentrations being equal (they are not the same as the bulk concentrations due to confinement).
In the OFF state, however, both cation and anion concentrations are small, with the concentration of the coion being much lower than the concentration of the counterion.
A gap appears between the two profiles.
This is what we mean by overlap of double layers and depletion of coions.
The degree of overlap can be characterized by the gap.

From the point of view of scaling, the relevant question is the degree of this depletion as a function of the $R_{\mathrm{P}}/\lambda$ parameter. 
That is shown by the two rightmost columns of Fig.\ \ref{fig2} for the OFF state in the negatively charged zone (circled by a green ellipse on the left). 
Profiles are shown for different values of $R_{\mathrm{P}}/\lambda$ from top to bottom for two different values of $R_{\mathrm{P}}$ (middle and right columns).

Figure \ref{fig2} shows that the degree of overlap (indicated by green arrows) depends on the parameter $R_{\mathrm{P}}/\lambda$.
The degree of overlap is larger (the degree of coion depletion is larger) if this parameter is smaller.
The value of $R_{\mathrm{P}}/\lambda$ can be small either if the pore radius is small, or the concentration is small ($\lambda$ is large).
If we compare the columns for $R_{\mathrm{P}}=1$ and $2$ nm (middle and right columns), it is apparent the the degree of overlap is similar in the two cases for a given $R_{\mathrm{P}}/\lambda$, but the concentrations are smaller for the larger $R_{\mathrm{P}}$ case.

The take-home message of Fig.\ \ref{fig2} is that the appearance of depletion zones is related to the $R_{\mathrm{P}}/\lambda$ ratio, so the rectification should also be a function of that parameter.

\begin{figure*}[t]
	\begin{center}
		\includegraphics*[width=0.9\textwidth]{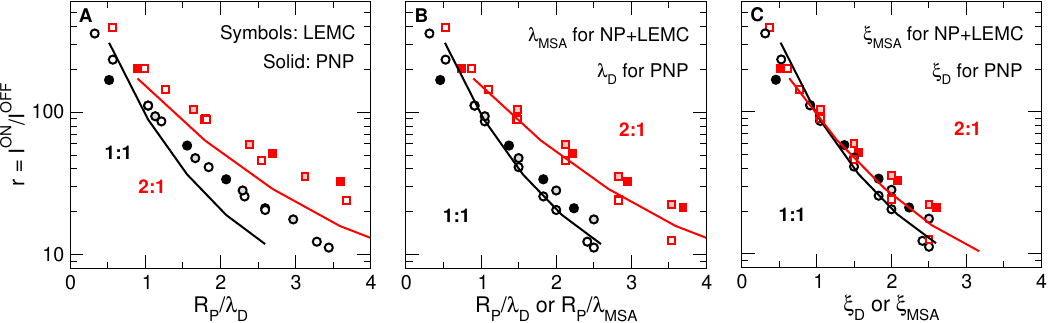}
		\vspace{0.5cm}
		\caption{ 
		Rectification defined as $r=I^{\mathrm{ON}}/I^{\mathrm{OFF}}$ as a function of various scaling parameters: (A) $R_{\mathrm{P}}/\lambda_{\mathrm{D}}$ for both LEMC and PNP, (B) $R_{\mathrm{P}}/\lambda_{\mathrm{D}}$ for PNP and  $R_{\mathrm{P}}/\lambda_{\mathrm{MSA}}$ for LEMC, and (C) $\xi_{\mathrm{D}}$ for PNP and $\xi_{\mathrm{MSA}}$ for LEMC, where $\xi_{D}=R_{\mathrm{P}}/(\lambda_{\mathrm{D}}z_{\mathrm{if}})$ and $\xi_{\mathrm{MSA}}=R_{\mathrm{P}}/(\lambda_{\mathrm{MSA}}z_{\mathrm{if}})$ with $z_{\mathrm{if}}=\sqrt{z_{+}|z_{-}|}$.
		Black and red symbols/lines refer to 1:1 and 2:1 systems, respectively.
		Symbols and lines refer to LEMC and PNP results, respectively.
		Filled symbols and lines have been obtained for a fixed concentration ($c=0.1$ M) with varying $R_{\mathrm{P}}$, symbols that are white inside have been obtained for fixed radii, $R_{\mathrm{P}}=1$ nm, with varying $c$, while symbols that are lighter colored inside have been obtained for fixed radii, $R_{\mathrm{P}}=2$ nm, with varying $c$.
		}
		\label{fig3}
	\end{center}
\end{figure*}

That is shown in Fig.\ \ref{fig3}A.
This panel shows rectification as a function of the $R_{\mathrm{P}}/\lambda_{\mathrm{D}}$ parameter for 1:1 and 2:1 electrolytes (3:1 and 2:2 electrolytes will be presented later to avoid clutter in this figure).
For the screening length, the Debye length $\lambda_{\mathrm{D}}$ is used in this panel as a first step.

In Fig.\ \ref{fig3}A scaling works in term of the $R_{\mathrm{P}}/\lambda_{\mathrm{D}}$ parameter for all the four cases separately (1:1 and 2:1 electrolytes computed with either NP+LEMC or PNP); the results are located along a smooth curve (e.g., all the different red symbols fall on a line, as do the black symbols).
The four curves, however, do not coincide.

Thus, there are basically two problems with Fig.\ \ref{fig3}A.
One is that the curves for NP+LEMC and PNP (symbols vs.\ lines) do not coincide.
Although there is nothing surprising in the fact that different methods using different degrees of approximations provide different results, we will show that it is possible to bring those results together from the point of view of scaling if we use an appropriate $\lambda$ for each of the different methods. 

The other problem is that the curves for 1:1 and 2:1 systems (black vs. red) do not coincide.
This means that rectification depends explicitly on $z_{+}$ and $z_{-}$.
Therefore, the $R_{\mathrm{P}}/\lambda_{\mathrm{D}}$ scaling works only in the $(R_{\mathrm{P}},c)$ parameter space.
We would, however, like to include the valences in the parameter set over which scaling is valid. 
This means that we need a new scaling parameter, specifically the one we defined in Eq.\ \ref{eq:xi}.

In the following two sections, we discuss and fix these problems.

\subsection{Screening length}
\label{sec:msa}

Let us deal with the question of how to compute the screening length, $\lambda$, first.
The Debye length is the result of a mean-field theory (PB) that fits the PNP theory, because the same degree of approximations are used in computing the $c_{i}(\mathbf{r})$ vs.\ \ $\mu_{i}(\mathbf{r})$ relation and in computing the screening length.

NP+LEMC, however, is a method that goes well beyond the mean-field level, so using $\lambda_{\mathrm{D}}$ in that case is questionable.
While we could squeeze some kind of screening length out of the simulation data directly, we have chosen a much simpler route.
We decided to use the screening length provided by the simplest and yet quite powerful statistical mechanical theory that can take into account ionic correlations (including finite ion sizes) via an integral equation (Orstein-Zernike) treatment.

The screening parameter from MSA is defined as  \cite{blum_mp_1975,blum_jcp_1977,2000_nonner_bj_1976}
\begin{equation}
 \lambda_{\mathrm{MSA}}=\dfrac{1}{2\Gamma},
\end{equation} 
where the $\Gamma$ is given by the implicit relation 
\begin{equation}
 4\Gamma^{2} = \frac{e^{2}}{\epsilon_{0}\epsilon kT} \sum_{i} c_{i} \left( \frac{z_{i}-\eta d_{i}^{2}}{1+\Gamma d_{i}} \right)^{2} ,
\end{equation} 
where $d_{i}=2R_{i}$ is the ionic diameter, 
\begin{equation}
 \eta = \frac{1}{\Omega} \frac{\pi}{2\Delta} \sum_{i} \frac{c_{i}d_{i}^{3}}{1+\Gamma d_{i}},
\end{equation} 
\begin{equation}
 \Omega = 1+ \frac{\pi}{2\Delta} \sum_{i} \frac{c_{i}d_{i}^{3}}{1+\Gamma d_{i}},
\end{equation} 
and 
\begin{equation}
 \Delta = 1- \frac{\pi}{6}\sum_{i}c_{i}d_{i}^{3}.
\end{equation} 
Note that for the case of $R_{+}=R_{-}$ considered here, these equations reduce to a simple quadratic equation with exactly one positive root.
Also, the MSA screening parameter is the Debye length in the limit of point ions:
\begin{equation}
 \lim_{d_{i}\rightarrow0} \lambda_{\mathrm{MSA}} = \lambda_{\mathrm{D}}.
\end{equation} 

If we rescale Fig.\ \ref{fig3}A and use $\lambda_{\mathrm{MSA}}$ for the LEMC points instead of $\lambda_{\mathrm{D}}$ (see Fig.\ \ref{fig3}B), we obtain a much better agreement between the LEMC and PNP data (the lines of PNP overlap the symbols of LEMC for 1:1 and for 2:1 electrolytes).
Since the PNP curves are unchanged, the LEMC data are shifted leftward by this rescaling, because MSA screening lengths for a given concentration are larger than the Debye lengths. 
This is valid for both 1:1 and 2:1 electrolytes.

From now on, when we talk about a general $\lambda$ value, we will mean the $\lambda_{\mathrm{MSA}}$ value in NP+LEMC and the $\lambda_{\mathrm{D}}$ value in PNP.

\subsection{Extension to multivalent ions}
\label{sec:multivalent}

Next, we turn to the problem that the curves for 1:1 and 2:1 electrolytes do not coincide in Fig.\ \ref{fig3}A and B.
As we stated in the introduction, this problem can be fixed by introducing the $\xi=R_{\mathrm{P}}/(\lambda z_{\mathrm{if}})$ parameter, where $\lambda$ is either $\lambda_{\mathrm{MSA}}$ or $\lambda_{\mathrm{D}}$ depending whether we use LEMC or PNP, and $z_{\mathrm{if}}=\sqrt{z_{+}|z_{-}|}$.

As seen in Fig.\ \ref{fig3}C, rescaling with the $z_{\mathrm{if}}$ parameter brings the 1:1 and 2:1 curves together.
As we will see later, it works for 3:1 and 2:2 electrolytes as well.
Before that, however, we provide an argument as to why the $z_{\mathrm{if}}$ factor works.

In previous work \cite{valisko_jml_2007} we studied the anomalous  temperature dependence of the capacitance of the electrical double layer for valence-asymmetric electrolytes.
The temperature was characterized by the reduced temperature
\begin{equation}
 T^{*}=\dfrac{4\pi \epsilon_{0}\epsilon kTd}{e^{2}} ,
 \label{eq:tred}
\end{equation} 
where $d$ was the diameter of ions (the same for anions and cations).
The reduced temperature is effectively the reciprocal of the strength of the interaction energy between two monovalent ions at contact (i.e., at a distance $d$ between the center charges) relative to $kT$.
We showed in that paper \cite{valisko_jml_2007} that capacitances behave the same way for 1:1, 2:1, 3:1, and 2:2 electrolytes if we plot them as functions of $T^{*}/|z_{+}z_{-}|$ instead of just $T^{*}$.

This scaling was confirmed by the theoretical work of di Caprio et al.\ \cite{dicaprio_mp_2006} whose field-theoretical approach was based on expressing the Hamiltonian as a functional of the charge density field, $q(\mathbf{r})=z_{+}c_{+}(\mathbf{r})+z_{-}c_{-}(\mathbf{r})$, and the total density field, $s(\mathbf{r})=c_{+}(\mathbf{r})+c_{-}(\mathbf{r})$.

The ion-ion (II) interaction in the Hamiltonian in the treatment of di Caprio et al.\ \cite{dicaprio_mp_2006} is given as
\begin{equation}
 \frac{H^{\mathrm{II}}[q(\mathbf{r})]}{kT}  = 
\frac{e^{2}}{4\pi\epsilon_{0}\epsilon kT}
\int \frac{q(\mathbf{r})q(\mathbf{r}')}{|\mathbf{r}-\mathbf{r}'|}d\mathbf{r} d\mathbf{r}'.
 \label{eq:coulomb1}
\end{equation} 
This functional can be written in the form 
\begin{equation}
 \frac{H^{\mathrm{II}}[q(\mathbf{r})]}{kT}  = 
\frac{1}{8\pi \bar{c}\lambda^{2}_{\mathrm{D}}z_{\mathrm{if}}^{2}}
\int \frac{q(\mathbf{r})q(\mathbf{r}')}{|\mathbf{r}-\mathbf{r}'|}d\mathbf{r} d\mathbf{r}',
 \label{eq:coulomb2}
\end{equation} 
if we write the Debye length as
\begin{equation}
\lambda_{\mathrm{D}}=\left(\frac{\bar{c}z^{2}_{\mathrm{if}}e^{2}}{\epsilon_{0}\epsilon kT} \right)^{-1/2},
\label{eq:debye-zif}
\end{equation} 
where $\bar{c}=c_{+}+c_{-}$ is the total density in the bulk.
To develop Eq.\ \ref{eq:debye-zif}, the relation
\begin{equation}
 z_{+}^{2}c_{+}+z_{-}^{2}c_{-} = z_{+}|z_{-}| \bar{c} = z_{\mathrm{if}}^{2} \bar{c}
\label{eq:zifc}
\end{equation} 
was used. 
This relates the ionic strength to $\bar{c}$.
It is through this relation that the parameter $z_{\mathrm{if}}$ appears.
The charge neutrality condition $z_{+}c_{+}-|z_{-}|c_{-}=0$ was used in the derivation of Eq.\ \ref{eq:zifc}.

Starting from Eq.\ \ref{eq:coulomb2}, di Caprio et al.\ \cite{dicaprio_mp_2006} introduced a formal scaling by defining a scaled density field as $q(\mathbf{r})\rightarrow Q(\mathbf{r})= q(\mathbf{r})/z_{\mathrm{if}}$ and a scaled unit charge as $e\rightarrow \tilde{e}=ez_{\mathrm{if}}$.
Doing that, Eq.\ \ref{eq:coulomb2} can be written in the form
\begin{equation}
 \frac{H^{\mathrm{II}}[q(\mathbf{r})]}{kT}  = 
\frac{1}{8\pi \bar{c}\lambda^{2}_{\mathrm{D}}}
\int \frac{Q(\mathbf{r})Q(\mathbf{r}')}{|\mathbf{r}-\mathbf{r}'|}d\mathbf{r} d\mathbf{r}',
 \label{eq:coulomb3}
\end{equation} 
where the Debye length is expressed as
\begin{equation}
\lambda_{\mathrm{D}}=
\left( \frac{\bar{c}\tilde{e}^{2}}{\epsilon_{0}\epsilon kT} \right)^{-1/2}
\label{eq:debye-zif2}
\end{equation} 
that is the same as for a 1:1 electrolyte but using the rescaled unit charge $\tilde{e}$ instead of $e$.
It was shown that with this rescaling the equations for charge-symmetric systems are partly recovered and the field theory using the rescaled densities provides good results for the electrical double in the case of charge-asymmetric electrolytes as well \cite{dicaprio_mp_2006}.

To justify our use of $z_{\mathrm{if}}$ in the $\xi$ parameter, we take another look at Eq.\ \ref{eq:coulomb2}, which is equivalent to Eq. \ref{eq:coulomb3}. 
This can be rewritten with the $z_{\mathrm{if}}$-modified Debye length
\begin{equation}
 \tilde{\lambda}_{\mathrm{D}} = \lambda_{\mathrm{D}} z_{\mathrm{if}} 
\end{equation} 
as
\begin{equation}
 \frac{H^{\mathrm{II}}[q(\mathbf{r})]}{kT}  = 
\frac{1}{8\pi \bar{c} \tilde{\lambda}^{2}_{\mathrm{D}}}
\int \frac{q(\mathbf{r})q(\mathbf{r}')}{|\mathbf{r}-\mathbf{r}'|}d\mathbf{r} d\mathbf{r}'
 \label{eq:coulomb4} 
\end{equation} 
which has a form invariant for electrolyte charge asymmetry; it is hidden in $\tilde{\lambda}_{\mathrm{D}}$ and $q(\mathbf{r})$.
This equation shows that the electrolyte's behavior is rather governed by the modified Debye length, $\tilde{\lambda}_{\mathrm{D}}$, instead of $\lambda_{\mathrm{D}}$.
This is in agreement with the definition of the parameter $\xi=R_{\mathrm{P}}/\tilde{\lambda}$.
In our application of this modified screening length, we have gone one step further, namely to use the screening length most appropriate for the system in question.

\section{Discussion}
\label{sec:discussion}

\subsection{Scaling and ion correlations}
\label{sec:discussion1}

\begin{figure}[t]
	\begin{center}
		\includegraphics*[width=0.42\textwidth]{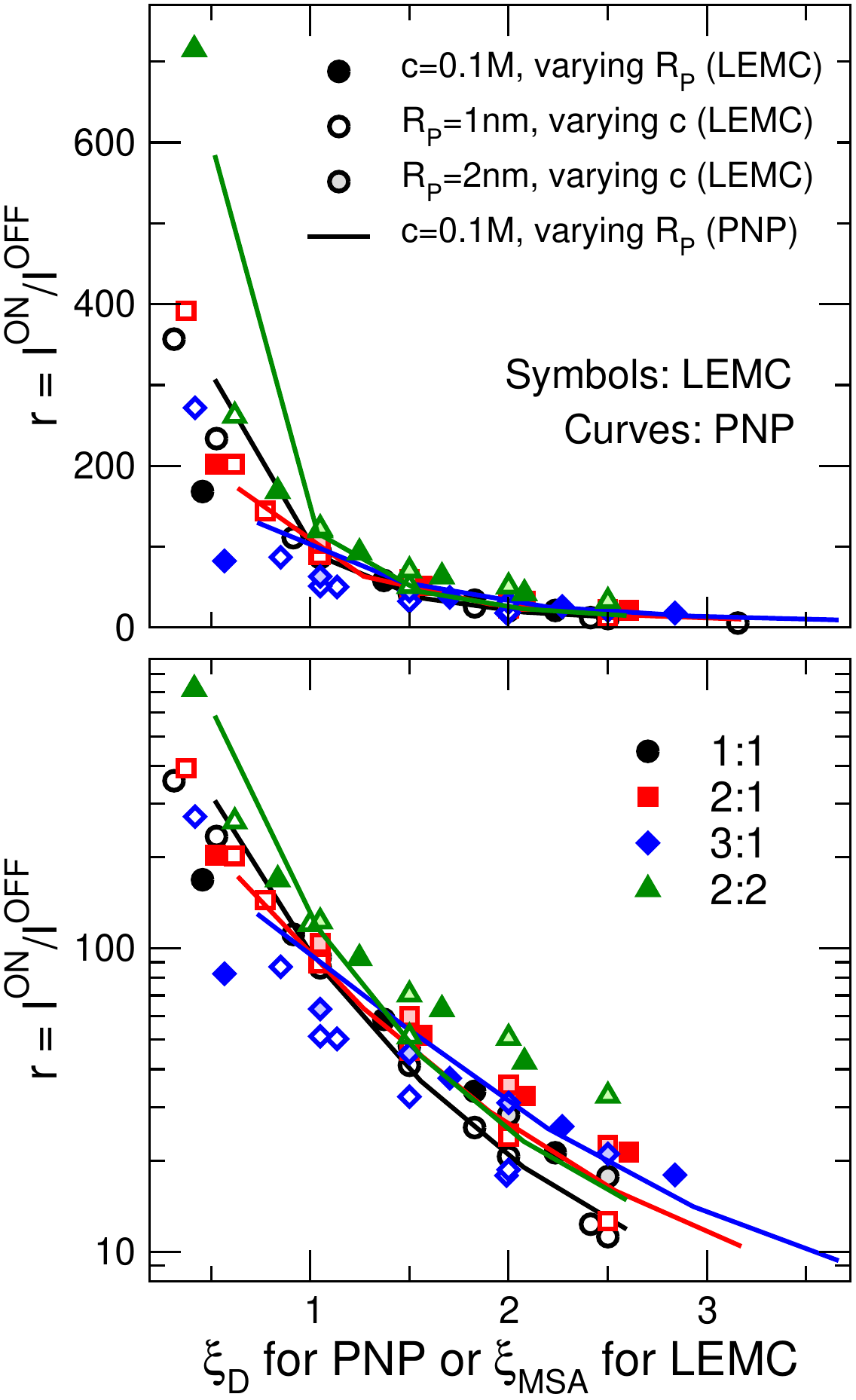}
		\vspace{0.5cm}
		\caption{ 
		Rectification defined as $r=I^{\mathrm{ON}}/I^{\mathrm{OFF}}$ as a function of the $\xi=R_{\mathrm{P}}/(\lambda z_{\mathrm{if}})$ parameter for various systems as indicated in the legend. 
		Black, red, blue, and green symbols/lines refer to 1:1, 2:1, 3:1, and 2:2 systems, respectively.
		Symbols and lines refer to LEMC and PNP results, respectively.
		Filled symbols and lines have been obtained for a fixed concentration ($c=0.1$ M) with varying $R_{\mathrm{P}}$, symbols that are white inside have been obtained for fixed radii, $R_{\mathrm{P}}=1$ nm, with varying $c$, while symbols that are lighter colored inside have been obtained for fixed radii, $R_{\mathrm{P}}=2$ nm, with varying $c$.
		The ordinates are plotted on a linear scale in the top panel, while on a logarithmic scale in the bottom panel.
 		The linear scale accentuates deviations at small $\xi$ values, while the logarithmic scale  accentuates deviations at large $\xi$ values. 
		}
		\label{fig4}
	\end{center}
\end{figure}

So far we have established that the $\xi$-scaling for rectification works for 1:1 and 2:1 electrolytes (Fig.\ \ref{fig3}C). 
Next, we compile this data and that for 3:1 and 2:2, where ion correlations are very strong, and show it in Fig.\ \ref{fig4}.
The top and bottom panels show the same results but on linear and logarithmic scales, respectively.
The linear scale enhances the deviations at small $\xi$ values, while the logarithmic scale rather enhances the deviations at large $\xi$ values.

The scaling is not perfect, but, taken collectively, all these disparate curves essentially fall on top of each other. 
This is remarkable for three reasons. 

\textit{First}, considering how easily they could be far apart with an inappropriate scaling parameter (as shown in Figs.\ \ref{fig3}A and B), Fig.\ \ref{fig4} shows that the right screening parameter plays a large role in this $r(\xi)$ curve. 
We suspect that part of the scatter in Fig.\ \ref{fig4} can be attributed to the fact that the MSA theory becomes less accurate as valence increases, especially as high as $+3$. 

\textit{Second}, it is remarkable that two simple tweaks to the usual $R_{\mathrm{P}}/\lambda_{\mathrm{D}}$ scaling can describe systems with vastly different complex ionic correlations. 
While Monte Carlo simulations naturally compute these correlations, great effort has been made by many theorists to develop sophisticated statistical mechanical theories that account for these correlations (e.g., density functional theories, integral equation theories with various closures).
Our results show that for pore rectification, these correlations can be taken into account (at least to first-order) with a better screening length $\lambda_{\mathrm{MSA}}$ and with $z_{\mathrm{if}}$. 
The importance of $z_{\mathrm{if}}$ cannot be overstated in making the scaling work. 
Figs.\ \ref{fig3}B and C show its effect and that just using a better screening length ($\lambda_{\mathrm{MSA}}$ instead of $\lambda_{\mathrm{D}}$) is insufficient. 
In fact, as di Caprio et al.\ wrote \cite{dicaprio_mp_2006}: ``This shows that this scaling related to the ionic strength parameter $z_{\mathrm{if}}$ can be considered as a primary effect.''

\textit{Third}, the details of the nanoscale physics inside the pore are very different with the different ionic correlations for 1:1, 2:1, 3:1, and 2:2 electrolytes.
Although it is obvious that different device-level behaviors emerge from these different molecular-level correlations, it is not at all obvious that this emergent behavior should be described by a simplistic scaling as a function of the $\xi$ variable.

If we want to shed a light on the mechanisms behind this, we can analyze the concentration profiles, $c_{i}(z,r)$, because they bridge the hard-to-quantify microscopic correlations and macroscopic observable quantities such as currents.
Currents are the integrals of the flux densities on the left hand side of the NP equation (Eq.\  \ref{eq:np}), while the profiles on the right hand side of the NP equation determine how currents behave in various conditions.
Although $c_{i}(z,r)$ profiles are available from the simulations, it is more practical to analyze the axial (radial) profiles that are averaged over the radial (axial) dimension, as we did in Fig.\ \ref{fig1} (Fig.\ \ref{fig2}). 

\begin{figure}[t]
	\begin{center}
		\includegraphics*[width=0.47\textwidth]{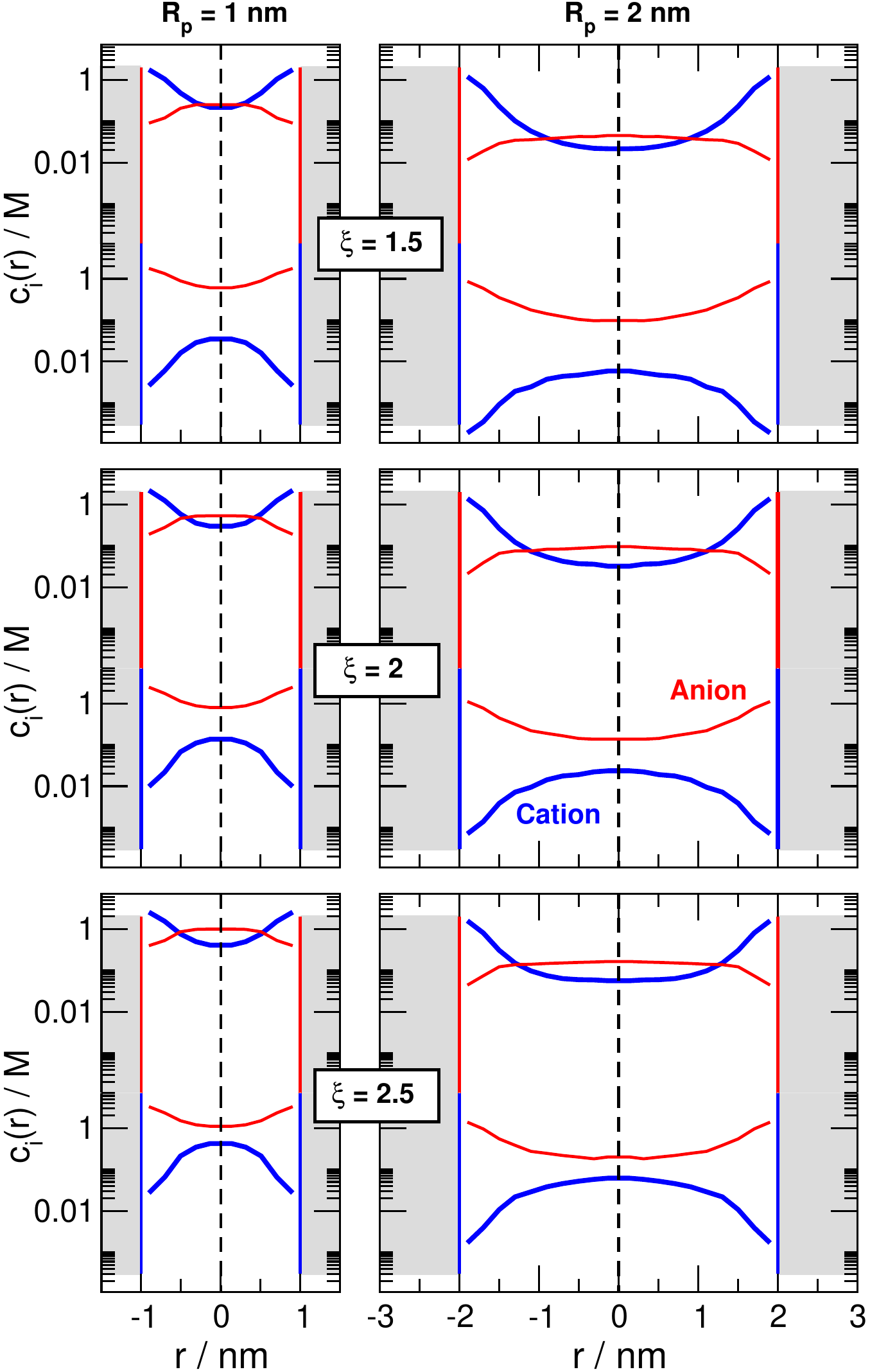}
		\vspace{0.5cm}
		\caption{
		The radial concentration profiles for 2:1 electrolytes in the OFF state for different values of the parameter $\xi=\xi_{\mathrm{MSA}}$.
		The profiles, obtained by averaging over the negative (indicated by vertical red lines) and positive (indicated by vertical blue lines) regions, have been computed by the NP+LEMC method.
		Results are shown for pore radii $R_{\mathrm{P}}=1$ nm (left column) and $R_{\mathrm{p}}=2$ nm (right column) for $\xi_{\mathrm{MSA}}=1.5$, $2$, and $2.5$ (from top to bottom).
        The concentrations that correspond to these state points can be found in Table S1 of the SI (they are in the $c=0.0466-0.905$ M concentration range).
		Thick blue lines refer to divalent cations, while thin red lines refer to monovalent anions. 
		}
		\label{fig5}
	\end{center}
\end{figure}

\begin{figure*}[t]
	\begin{center}
		\includegraphics*[width=0.99\textwidth]{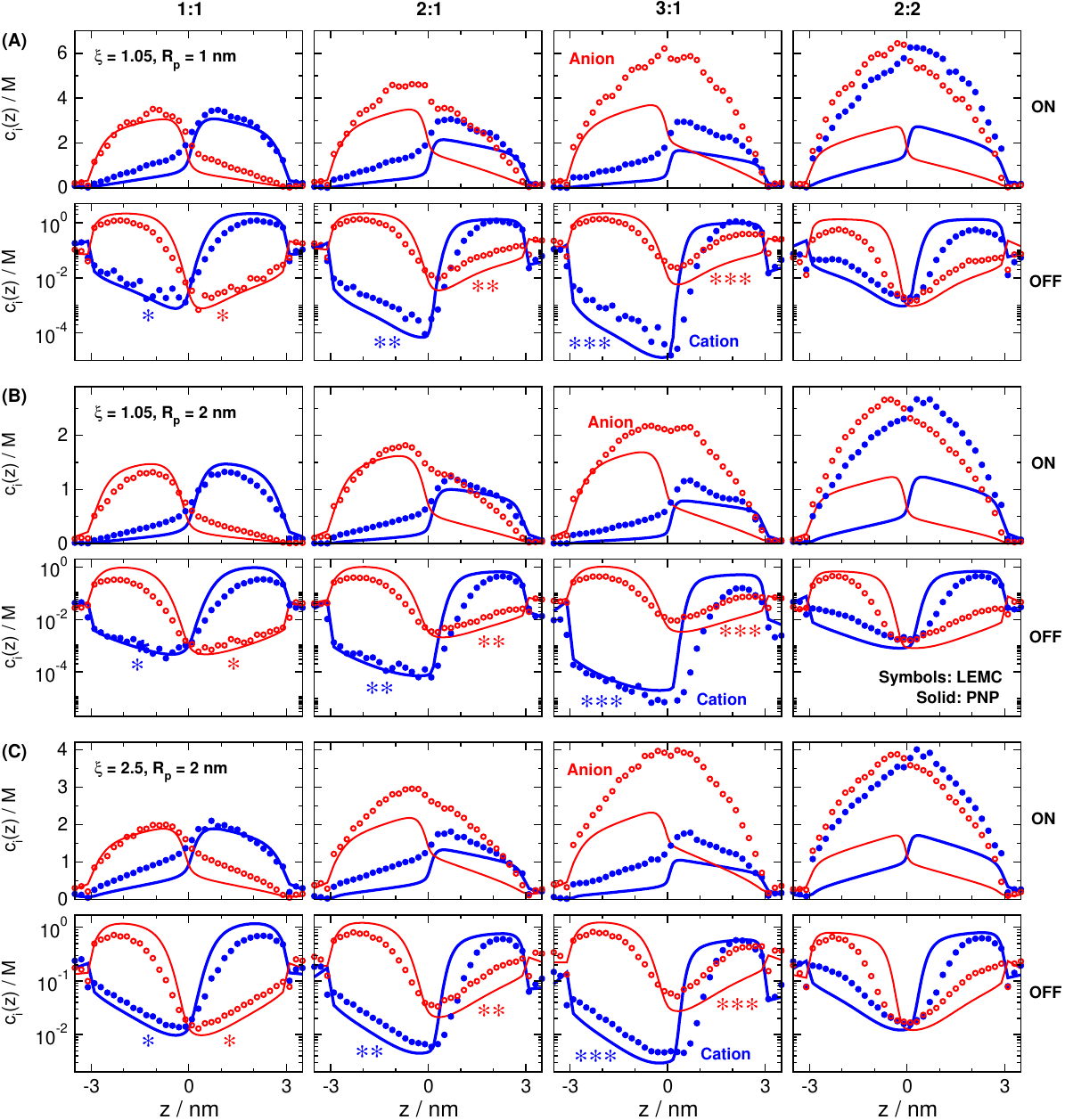}
		\vspace{0.5cm}
		\caption{
		Axial concentration profiles for 
		(A) $\xi=1.05$, $R_{\mathrm{P}}=1$ nm, 
		(B) $\xi=1.05$, $R_{\mathrm{P}}=2$ nm, and 
		(C) $\xi=2.5$, $R_{\mathrm{P}}=2$ nm.
		Columns refer to 1:1, 2:1, 3:1, and 2:2 electrolytes as indicated.
		Top and bottom rows in a panel refer to the ON and OFF states, respectively.
		Blue filled and red open symbols refer to cation and anion profiles, respectively, as obtained by NP+LEMC.
		Thick blue and thin red lines refer to cation and anion profiles, respectively, as obtained by PNP.
		LEMC and PNP results refer to different concentrations (see Table S1), because they refer to screening length values computed differently ($\lambda_{\mathrm{D}}$ vs.\ $\lambda_{\mathrm{MSA}}$).
		Note that the ordinates are plotted on a linear scale in the ON state and on a logarithmic scale in the OFF state.
		}
		\label{fig6}
	\end{center}
\end{figure*}

Figures \ref{fig1} and \ref{fig2} show that there is an interesting coupling between the radial and axial dimensions.
Fig.\ \ref{fig2} illustrates for a 1:1 electrolyte that the radial dimension determines how the double layers behave and how the depletion zones are formed.
The device behavior, however, is determined by how those depletion zones appear along the ionic pathway (i.e., in the axial dimension) \cite{hato_pccp_2017}.
Rectification is determined by the voltage-sensitive formation of the depletion zones as shown by the axial concentration profiles in Fig.\ \ref{fig1}.
This coupling is also present for multivalent ions and is analyzed in Figs.\ \ref{fig5} and \ref{fig6}.

\vspace{0.3cm}
Figure \ref{fig5} is the counterpart of Fig.\ \ref{fig2} for 2:1 electrolytes.
We plot radial concentration profiles in the OFF state to analyze how double layers overlap and coion depletion zones are formed.
If the divalent cations (blue lines) are the counterions, they are attracted to the oppositely charged wall (vertical red lines) strongly.
The divalent cations, however, attract more anions (red lines) into the pore due to stronger correlations between them, even if the anions are repulsed by the pore charge, a phenomenon that eventually leads to charge inversion \cite{he_jacs_2009,garciagimenez_pre_2010,gurnev_langmuir_2012,li_nl_2015,ramirez_jmembsci_2018,mashayak_jcp_2018,chou_nl_2018,voukadinova_jcp_2019}.
The depletion zones of anions, therefore, are not so deep.

If the monovalent anions are the counterions, on the other hand, the situation is reversed.
The divalent cations (the coions) are repulsed by the positive pore charge  (vertical blue lines) strongly.
This is the dominant effect, so the divalent cations form very deep depletion zones when they are the coions.

What is remarkable is how this asymmetric behavior scales with $\xi$.
The gaps between the cation and anion profiles (that roughly characterize double layer overlap) change with $\xi$ (from top to bottom) showing a clear tendency: the gap decreases (degree of overlap decreases) as $\xi$ increases.
This tendency is the same for the two pore radii shown (compare left and right columns).

\vspace{0.3cm}
Figure \ref{fig6} shows how these effects in the radial dimension manifest themselves in the axial dimension.
It shows all the axial concentration profiles for three selected sets of $\xi$ and $R_{\mathrm{P}}$ (panels A-C) both in the ON and OFF states as obtained by both methods.

The fact (observed and discussed above for Fig.\ \ref{fig5}) that the depletion zones of multivalent cations are deeper in the OFF state is even more apparent in Fig.\ \ref{fig6}: the depletion zones get deeper as cation valences increase from $+1$ to $+3$; see blue symbols in the second row from left to right in the first three columns (1:1, 2:1, and 3:1; note the logarithmic scale of concentration).
For the anions, the opposite trend is observed.
Both trends are even more clearly visible in Fig.\ S1 of the Supporting Information (SI), where we plot the 1:1, 2:1, and 3:1 cases in one panel.
To make sure that these trends are visible in Fig.\ \ref{fig6}, we indicated these cases with $\ast$, $\ast\ast$, and ${\ast}{\ast}{\ast}$ symbols, respectively, in blue for cations and in red for anions.

In the ON state, concentrations are larger in the pore, so the effect of strong ionic correlations in the multivalent cases (2:1, 3:1, and 2:2) is more apparent.
In PNP, the anion profiles do not change much as $z_{+}$ increases (from 1:1 to 3:1), while the cation profiles decrease, because the multivalent cations provide more charge to balance the pore charge.
This is the natural outcome of the mean-field PNP theory.
In contrast, the LEMC cation profiles do not change much and the anion profiles increase as $z_{+}$ increases.
This is the result of the ionic correlations between cations and anions that are properly computed by LEMC.
The multivalent cations drag the anions along with them into the pore.
Also, this trend is more visible in Fig.\ S1.

The effect of strong ionic correlations can also be seen by comparing the 2:2 case to the 1:1 case.
Both cation and anion profiles are elevated and a much larger density electrolyte is formed inside the pore than that implied by the mean-field PNP theory.
The divalent cations and divalent anions associate so strongly that they drag each other along into the pore.
This phenomenon is absent in PNP.
This effect is even more visible in Fig.\ S2 of the SI, where we plot the 1:1 and 2:2 cases in one panel.

If we compare panels A and B (they refer to the same $\xi$, but different $R_{\mathrm{P}}$ values), we observe quantitatively similar behavior of the OFF-state profiles; note that the ordinates of panels A and B show similar concentration ranges.
If we compare panels B and C (they refer to the same $R_{\mathrm{P}}$, but different $\xi$ values), we observe quantitatively different behavior of the OFF-state profiles;  note that the ordinates of panels B and C show very different concentration ranges.

\afterpage{\clearpage}

\begin{figure*}[t!]
	\begin{center}
		\includegraphics*[width=0.6\textwidth]{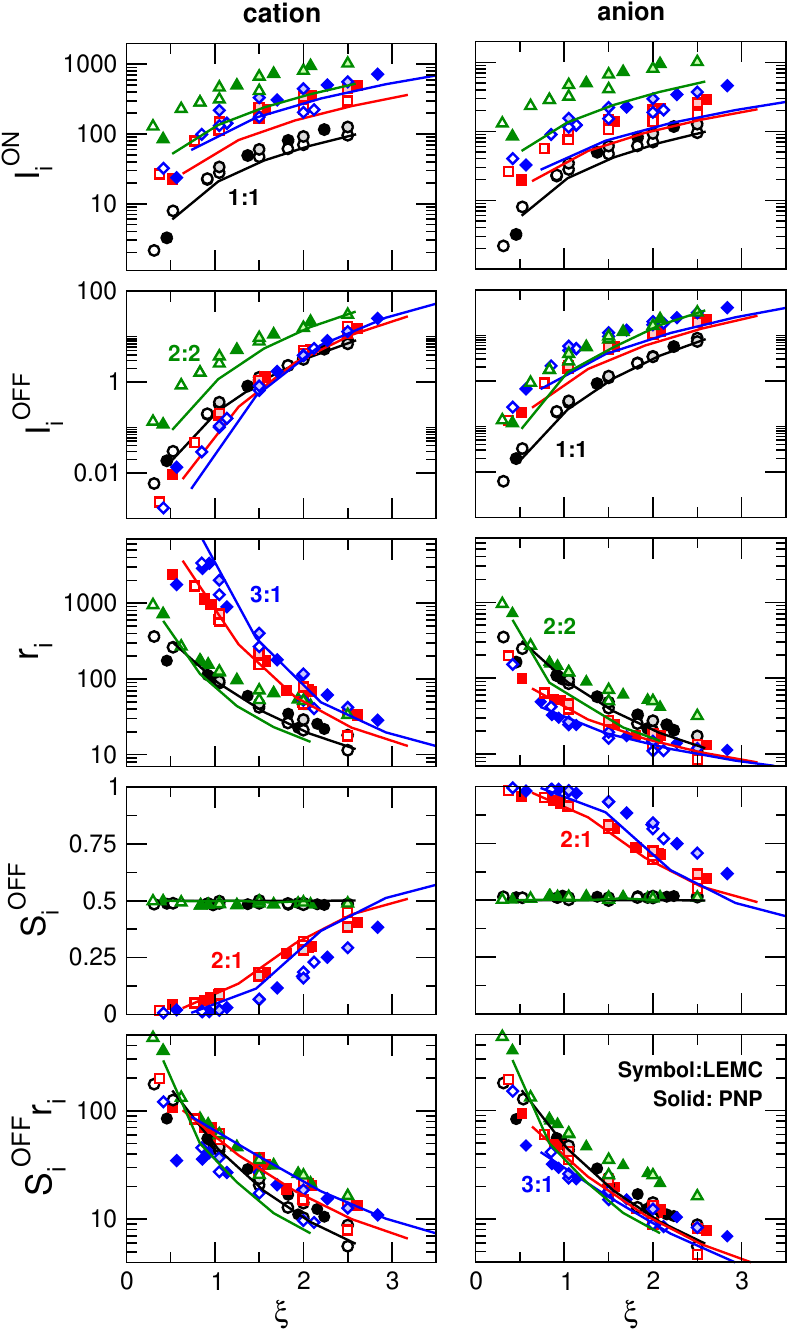}
		\vspace{0.5cm}
		\caption{ 
		Analysis of the relation of rectification and selectivity on the basis of equation $r=S_{+}^{\mathrm{OFF}}r_{+}+S_{-}^{\mathrm{OFF}}r_{-}$, where $S_{i}^{\mathrm{OFF}}=I_{i}^{\mathrm{OFF}}/I^{\mathrm{OFF}}$ is the selectivity for ionic species $i$ in the OFF state and $r_{i}=I_{i}^{\mathrm{ON}}/I_{i}^{\mathrm{OFF}}$ is the rectification for the currents carried by ionic species $i$.
		Panels from top to bottom show ionic currents in the ON state ($I_{i}^{\mathrm{ON}}$), in the OFF state ($I_{i}^{\mathrm{OFF}}$), their ratio ($r_{i}$), OFF-state selectivity ($S_{i}^{\mathrm{OFF}}$), and the product of the latter two ($S_{i}^{\mathrm{OFF}}r_{i}$).
		Left and right column refer to cations and anions, respectively.
		Colors, symbols, and lines have the same meaning as in Fig.\ \ref{fig4}.
	}
		\label{fig7}
	\end{center}
\end{figure*}

\afterpage{\clearpage}

It is important to emphasize that Fig.\ \ref{fig6} is not a \textit{direct} comparison between LEMC and PNP, because the two methods refer to different concentrations (see Table S1).
The purpose of Fig.\ \ref{fig6} is to show trends as functions of $z_{+}$ at fixed $\xi$ values.
If we plot the concentration profiles normalized by the bulk concentrations, $c_{i}(z)/c_{i}$, the agreement between the LEMC and PNP data is much better (see Fig.\ S3 in the SI).

\subsection{Scaling and selectivity}
\label{sec:individual}

So far, we showed results for rectification expressed in terms of the total current (Eq.\ \ref{eq:totrect}, Figs.\ \ref{fig3} and \ref{fig4}), because that is the primary measurable device function. 
At the same time, we showed concentrations of individual ionic species (Figs.\ \ref{fig1}, \ref{fig2}, \ref{fig5}, and \ref{fig6}).
In the following, we show how the behavior for individual ionic species add up to the measurable overall behavior (total currents and their rectification).

We define the rectification of ionic species $i$ as
\begin{equation}
 r_{i}=\dfrac{I_{i}^{\mathrm{ON}}}{I_{i}^{\mathrm{OFF}}}, 
\end{equation} 
where $I_{i}^{\mathrm{ON}}$ and $I_{i}^{\mathrm{OFF}}$ are the absolute values of currents carried by ionic species $i$ in the ON and OFF states, respectively.
If we express $r$ in terms of $r_{i}$ as
\begin{eqnarray}
 r&=&\dfrac{I^{\mathrm{ON}}}{I^{\mathrm{OFF}}}=\dfrac{I_{+}^{\mathrm{ON}}+I_{-}^{\mathrm{ON}}}{I^{\mathrm{OFF}}}
 = \dfrac{I_{+}^{\mathrm{ON}}}{I^{\mathrm{OFF}}}
 + \dfrac{I_{-}^{\mathrm{ON}}}{I^{\mathrm{OFF}}} \nonumber \\
 &=& \dfrac{I_{+}^{\mathrm{OFF}}}{I^{\mathrm{OFF}}} 
     \cdot
     \dfrac{I_{+}^{\mathrm{ON}}}{I_{+}^{\mathrm{OFF}}}
 +   \dfrac{I_{-}^{\mathrm{OFF}}}{I^{\mathrm{OFF}}}
     \cdot
     \dfrac{I_{-}^{\mathrm{ON}}}{I_{-}^{\mathrm{OFF}}} \nonumber \\
 &=& S_{+}^{\mathrm{OFF}} r_{+}  + S_{-}^{\mathrm{OFF}} r_{-} ,  
\label{eq:rec-sel}
 \end{eqnarray} 
we can see that rectification for the total current is a weighted sum of the rectifications for the individual ions weighted by the selectivities in the OFF state defined as
\begin{equation}
 S^{\mathrm{OFF}}_{i}=\dfrac{I_{i}^{\mathrm{OFF}}}{I^{\mathrm{OFF}}}
\end{equation} 
expressing the share of ionic species $i$ from the total current.
If $S_{i}^{\mathrm{OFF}}=1$, the pore is selective for ionic species $i$, if $S_{i}^{\mathrm{OFF}}=0$, the pore is selective for the other species, while if $S_{i}^{\mathrm{OFF}}=0.5$, the pore is non-selective.

Figure \ref{fig7} analyses all the quantities that appear in Eq.\ \ref{eq:rec-sel}.
From top to bottom, we plot the ionic currents in the ON state ($I_{i}^{\mathrm{ON}}$), in the OFF state ($I_{i}^{\mathrm{OFF}}$), their ratio ($r_{i}$), OFF-state selectivities ($S_{i}^{\mathrm{OFF}}$), and the products of the previous two ($S_{i}^{\mathrm{OFF}}r_{i}$).

The two top rows show that individual currents ($I_{i}^{\mathrm{ON}}$ and $I_{i}^{\mathrm{OFF}}$) have very different magnitudes for different electrolytes (beware the logarithmic scale).
Currents are different for the cations and the anions (for the 2:1 and 3:1 systems) both in the ON and OFF states.
This shows that the pore is selective in the case of valence-asymmetric electrolytes.

The rows for the individual rectification and OFF-state selectivity show that scaling does not work for these quantities.
The $r_{i}$ values for the 2:1 and 3:1 systems deviate from the 1:1 data; they are larger for cations and smaller for anions.
The OFF-state selectivity, however, shows the opposite trend.
If we take their product, however, the agreement is much better (bottom row).

This can be understood if we look at the 3:1 case (blue symbols and curves).
Rectification is large for the cation because the depletion zones of the trivalent cations is very deep, so the OFF-state current of the cations is very low.
At the same time, rectification is very small for the anion due to anion leakage (see the large $I_{-}^{\mathrm{OFF}}$ values in the second row).
Anion leakage is due to the fact that the depletion zones of anions are not very deep because of the strong correlations between the trivalent cations and the anions.
The trivalent cations, so to speak, bring the strongly correlated anions with them into the negative zones that otherwise repulse the anions \cite{voukadinova_jcp_2019}.

These large differences in individual rectifications, however, are balanced by selectivities.
The bipolar nanopore is selective for the anions for a charge-asymmetric electrolyte in the OFF state (see the $S_{i}^{\mathrm{OFF}}$ values in Fig.\ \ref{fig7}).
That is because the cations have much deeper depletion zones, so their OFF-state currents are much smaller.
The large rectification of the cation, $r_{+}$, therefore, contributes to the total rectification with a smaller weight as shown by Eq.\ \ref{eq:rec-sel}.

Scaling for the $S_{i}^{\mathrm{OFF}}r_{i}$ product works quite well for the 1:1 and 2:1 cases, while deviations appear for the 2:2 and 3:1 cases, where ionic correlations are stronger.

\begin{figure}[t]
	\begin{center}
		\includegraphics*[width=0.48\textwidth]{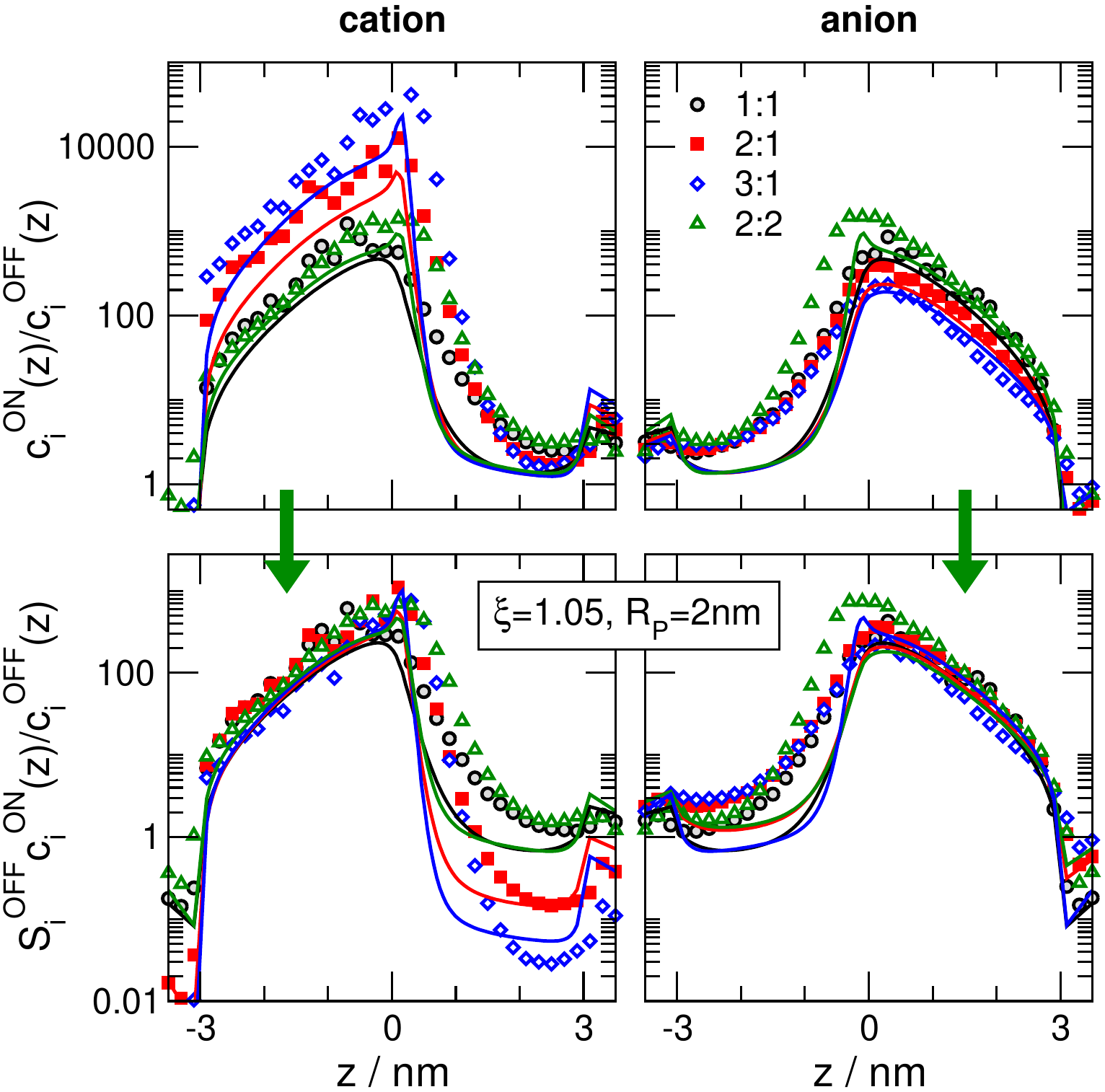}
		\vspace{0.5cm}
		\caption{  
		The $c_{i}^{\mathrm{ON}}(z)/c_{i}^{\mathrm{OFF}}(z)$ ratio for $\xi=1.05$ and $R_{\mathrm{P}}=2$ nm for 1:1, 2:1, 3:1, and 2:2 electrolytes (top row).
		Colors have the same meaning as in Fig.\ \ref{fig4}.
		Symbols and curves refer to LEMC and PNP results, respectively, both obtained for their own respective $\xi$ parameters.
		The $S_{i}^{\mathrm{OFF}}c_{i}^{\mathrm{ON}}(z)/c_{i}^{\mathrm{OFF}}(z)$ ratios are shown in the bottom row.  
		}
		\label{fig8}
	\end{center}
\end{figure}

A different way to see this is to divide the NP equation for the ON state with the NP equation for the OFF state. 
To first-order, the left-hand side result is $r_{i}$ since the area inside the pore is constant and so is the total flux.
On the right-hand side, the quantities that are largely different in the ON and OFF states are the concentration profiles since the $D_{i}(z)$ profiles are identical in this study.
Also, the $\mu_{i}(z)$ profiles are very similar because in absolute values they are the same in the left and right baths, and thus their variance is limited by this constraint. 
The concentrations, however, exhibit hugely different behavior in the ON and OFF states, see Figs.\ \ref{fig1} and \ref{fig6}.

In accordance with Eq.\ \ref{eq:rec-sel}, we  can expect that scaling works for the $c_{i}^{\mathrm{ON}}(z)/c_{i}^{\mathrm{OFF}}(z)$ ratio if we multiply it by $S_{i}^{\mathrm{OFF}}$.
This is shown by Fig.\ \ref{fig8}.
The top row of this figure shows the $c_{i}^{\mathrm{ON}}(z)/c_{i}^{\mathrm{OFF}}(z)$ profiles for a given $\xi$ and $R_{\mathrm{P}}$ for various electrolytes.
The curves depart especially for the cation.
If we multiply by $S_{i}^{\mathrm{OFF}}$, however, the curves line up especially in the depletion zone which is our main interest (bottom row).
In this figure (as in Fig.\ \ref{fig1}C), large peaks represent regions that contribute to rectification in the resistors connected  in series model.

\section{Conclusions}

Scaling is an important property in nature because it helps relate certain phenomena to many parameters in a simple way, often related to varying length and time scales \cite{west2017scale}.
In the world of nanodevices, scaling behavior for a device function (rectification, in this study) makes design of nanodevices easier. It may also help us understand the physics of the device function.

Here, we showed that rectification scales with $\xi$, where $\xi$ is a function of parameters $R_{\mathrm{P}}$, $c$, $R_{+}$,  $z_{+}$, $R_{-}$, and $z_{-}$.
The system's behavior can be described by a single parameter, $\xi$, thus the problem is seemingly reduced to a one dimensional one, provided that all the other parameters (e.g., pore length, pore charge) are kept fixed.
This is possible because there is a coupling between the radial dimension (i.e., in the cross-section) and the longitudinal dimension (i.e., down the axis of the pore) via the double layer overlap and the deepness of the depletion zones. 
Thus, scaling stems from a behavior in the radial dimension that determines the behavior in the longitudinal dimension, and, thus, device properties.

The concept of scaling can be paralleled with the idea of reduced quantities such as the reduced temperature defined in Eq.\ \ref{eq:tred}. 
These are dimensionless parameters that characterize many of the statistical mechanical properties of a system. 
A reduced density, $\rho^{*}=\rho/ d^{3}$, for example, can also be considered a scaling parameter in the sense that the system's properties, expressed in reduced (e.g., normalized or relative) quantities, depend on $\rho^{*}$ whatever is the number density $\rho$ or the particle diameter $d$. What matters is their ratio.

We showed results using two different methods that include two highly different degrees of approximations.
LEMC is a particle simulation method that includes ionic correlations (including finite sizes of ions). 
PNP, on the other hand, is a continuum theory that works on a mean-field level because it employs the PB theory.

We showed that the two methods can produce qualitatively the same scaling if we use the appropriate screening lengths that mirrors the physics in each model ($\lambda_{\mathrm{D}}$ for PNP because both are based on PB theory and $\lambda_{\mathrm{MSA}}$ for LEMC because both have correlated hard-sphere ions).
We were able to define a parameter based on a modified screening length ($\lambda\rightarrow \lambda z_{\mathrm{if}}$) that produced very similar scaling behavior for very different electrolytes from 1:1 to 3:1 to 2:2. Exactly why the rectification scales like this will require more work to understand, despite the theoretical results of di Caprio et al.\ \cite{dicaprio_mp_2006} which shed some light on the mechanisms behind it.

\subsection*{Supporting Information}

Table S1 contains the parameters of all the simulated state points. Figures S1 and S2 focus on details of Fig.\ \ref{fig6} to assist discussion. Figure S3 is an alternative of Fig.\ \ref{fig6} showing normalized concentration profiles.

\section*{Acknowledgements}
\label{sec:ack}

We gratefully acknowledge  the financial support of the National Research, Development and Innovation Office -- NKFIH K124353. 
BM acknowledges financial support from the Austrian Academy of Sciences \"OAW via the New Frontiers Grant NST-001.


\begin{thebibliography}{10}

\bibitem{daiguji_nl_2005}
H.~Daiguji, Y.~Oka, and K.~Shirono.
\newblock Nanofluidic diode and bipolar transistor.
\newblock {\em Nano Lett.}, 5(11):2274--2280, 2005.

\bibitem{vlassiouk_nl_2007}
I.~Vlassiouk and Z.~S. Siwy.
\newblock Nanofluidic diode.
\newblock {\em Nano Lett.}, 7(3):552--556, 2007.

\bibitem{yan_nl_2009}
R.~Yan, W.~Liang, R.~Fan, and P.~Yang.
\newblock Nanofluidic diodes based on nanotube heterojunctions.
\newblock {\em Nano Lett.}, 9(11):3820--3825, 2009.

\bibitem{albrecht_chapter_2013}
T.~Albrecht, T.~Gibb, and P.~Nuttall.
\newblock Ion transport in nanopores.
\newblock In {\em Engineered Nanopores for Bioanalytical Applications}, pages
  1--30. Elsevier {BV}, 2013.

\bibitem{Abgrall_2008}
P.~Abgrall and N.~T. Nguyen.
\newblock Nanofluidic devices and their applications.
\newblock {\em Anal. Chem.}, 80(7):2326--2341, 2008.

\bibitem{bocquet_csr_2010}
L.~Bocquet and E.~Charlaix.
\newblock Nanofluidics, from bulk to interfaces.
\newblock {\em Chem. Soc. Rev.}, 39(3):1073--1095, 2010.

\bibitem{daiguji_csr_2010}
H.~Daiguji.
\newblock Ion transport in nanofluidic channels.
\newblock {\em Chem. Soc. Rev.}, 39(3):901--911, 2010.

\bibitem{eijkel_csr_2010}
J.~C.~T. Eijkel and A.~van~den Berg.
\newblock Nanofluidics and the chemical potential applied to solvent and solute
  transport.
\newblock {\em Chem. Soc. Rev.}, 39(3):957, 2010.

\bibitem{zangle_csr_2010}
T.~A. Zangle, A.~Mani, and J.~G. Santiago.
\newblock Theory and experiments of concentration polarization and ion focusing
  at microchannel and nanochannel interfaces.
\newblock {\em Chem. Soc. Rev.}, 39(3):1014, 2010.

\bibitem{dal_cengio_jcp_2019}
S.~Dal Cengio and I.~Pagonabarraga.
\newblock Confinement-controlled rectification in a geometric nanofluidic
  diode.
\newblock {\em J. Chem. Phys.}, 151(4):044707, 2019.

\bibitem{dicaprio_mp_2006}
D.~Di Caprio, M.~Valisk{\'o}, M.~Holovko, and D.~Boda.
\newblock Anomalous temperature dependence of the differential capacitance in
  valence asymmetric electrolytes. {Comparison} of {Monte} {Carlo} simulation
  results and the field theoretical approach.
\newblock {\em Mol. Phys.}, 104(22-24):3777--3786, 2006.

\bibitem{constantin_pre_2007}
D.~Constantin and Z.~S. Siwy.
\newblock {Poisson-Nernst-Planck} model of ion current rectification through a
  nanofluidic diode.
\newblock {\em Phys. Rev. E}, 76(4):041202, 2007.

\bibitem{wolfram_jpcm_2010}
M.-T. Wolfram, M.~Burger, and Z.~S. Siwy.
\newblock Mathematical modeling and simulation of nanopore blocking by
  precipitation.
\newblock {\em J. Phys.-cond. Matt.}, 22(45):454101, 2010.

\bibitem{Ali_ACSnano_2012}
M.~Ali, S.~Nasir, P.~Ramirez, J.~Cervera, S.~Mafe, and W.~Ensinger.
\newblock Calcium binding and ionic conduction in single conical nanopores with
  polyacid chains: Model and experiments.
\newblock {\em {ACS} Nano}, 6(10):9247--9257, 2012.

\bibitem{gamble_jpcc_2014}
T.~Gamble, K.~Decker, T.~S Plett, M.~Pevarnik, J.-F. Pietschmann, I.~V.
  Vlassiouk, A.~Aksimentiev, and Z.~S. Siwy.
\newblock Rectification of ion current in nanopores depends on the type of
  monovalent cations -- experiments and modeling.
\newblock {\em J. Phys. Chem. C}, 118(18):9809--9819, 2014.

\bibitem{nikolaev_jce_2014}
A.~Nikolaev and M.~E. Gracheva.
\newblock {Poisson-Nernst-Planck} model for an ionic transistor based on a
  semiconductor membrane.
\newblock {\em J. Comput. Electron.}, 13(4):818--825, 2014.

\bibitem{hato_pccp_2017}
Z.~Hat{\'{o}}, M.~Valisk{\'{o}}, T.~Krist{\'{o}}f, D.~Gillespie, and D.~Boda.
\newblock Multiscale modeling of a rectifying bipolar nanopore: Explicit-water
  versus implicit-water simulations.
\newblock {\em Phys. Chem. Chem. Phys.}, 19(27):17816--17826, 2017.

\bibitem{valisko_jcp_2019}
M.~Valisk\'o, B.~Matejczyk, Z.~Hat\'o, T.~Krist\'of, E.~M\'adai, D.~Fertig,
  D.~Gillespie, and D.~Boda.
\newblock Multiscale analysis of the effect of surface charge pattern on a
  nanopore's rectification and selectivity properties: From all-atom model to
  {Poisson-Nernst-Planck}.
\newblock {\em J. Chem. Phys.}, 150(14):144703, 2019.

\bibitem{madai_pccp_2018}
E.~M\'adai, B.~Matejczyk, A.~Dallos, M.~Valisk\'o, and D.~Boda.
\newblock Controlling ion transport through nanopores: Modeling transistor
  behavior.
\newblock {\em Phys. Chem. Chem. Phys.}, 20(37):24156--24167, 2018.

\bibitem{he_jacs_2009}
Y.~He, D.~Gillespie, D.~Boda, I.~Vlassiouk, R.~S. Eisenberg, and Z.~S. Siwy.
\newblock Tuning transport properties of nanofluidic devices with local charge
  inversion.
\newblock {\em J. Am. Chem. Soc.}, 131(14):5194--5202, 2009.

\bibitem{garciagimenez_pre_2010}
E.~Garc{\'i}a-Gim{\'e}nez, A.~Alcaraz, and V.~M. Aguilella.
\newblock Overcharging below the nanoscale: {Multivalent} cations reverse the
  ion selectivity of a biological channel.
\newblock {\em Phys. Rev. E}, 81(2):021912, 2010.

\bibitem{gurnev_langmuir_2012}
P.~A. Gurnev and S.~M. Bezrukov.
\newblock Inversion of membrane surface charge by trivalent cations probed with
  a cation-selective channel.
\newblock {\em Langmuir}, 28(45):15824--15830, 2012.

\bibitem{li_nl_2015}
S.~X. Li, W.~Guan, B.~Weiner, and M.~A. Reed.
\newblock Direct observation of charge inversion in divalent nanofluidic
  devices.
\newblock {\em Nano Lett.}, 15(8):5046--5051, 2015.

\bibitem{ramirez_jmembsci_2018}
P.~Ramirez, J.~A. Manzanares, J.~Cervera, V.~Gomez, M.~Ali, I.~Pause,
  W.~Ensinger, and S.~Mafe.
\newblock Nanopore charge inversion and current-voltage curves in mixtures of
  asymmetric electrolytes.
\newblock {\em J. Membr. Sci.}, 563:633--642, 2018.

\bibitem{mashayak_jcp_2018}
S.~Y. Mashayak and N.~R. Aluru.
\newblock A multiscale model for charge inversion in electric double layers.
\newblock {\em J. Chem. Phys.}, 148(21):214102, 2018.

\bibitem{chou_nl_2018}
K.-H. Chou, C.~McCallum, D.~Gillespie, and S.~Pennathur.
\newblock An experimental approach to systematically probe charge inversion in
  nanofluidic channels.
\newblock {\em Nano Lett.}, 18(2):1191--1195, 2018.

\bibitem{voukadinova_jcp_2019}
A.~Voukadinova and D.~Gillespie.
\newblock Energetics of counterion adsorption in the electrical double layer.
\newblock {\em J. Chem. Phys.}, 150(15):154706, 2019.

\bibitem{kuo_langmuir_2001}
T.-C. Kuo, L.~A. Sloan, J.~V. Sweedler, and P.~W. Bohn.
\newblock Manipulating molecular transport through nanoporous membranes by
  control of electrokinetic flow:~ effect of surface charge density and {Debye}
  length.
\newblock {\em Langmuir}, 17(20):6298--6303, 2001.

\bibitem{ho_pnas_2005}
C.~Ho, R.~Qiao, J.~B. Heng, A.~Chatterjee, R.~J. Timp, N.~R. Aluru, and
  G.~Timp.
\newblock Electrolytic transport through a synthetic nanometer-diameter pore.
\newblock {\em Proc. Nat. Acc Sci.}, 102(30):10445--10450, 2005.

\bibitem{albesa_jmm_2013}
A.~G. Albesa, M.~Rafti, and J.~L. Vicente.
\newblock Trivalent cations switch the selectivity in nanopores.
\newblock {\em J. Mol. Model.}, 19(6):2183--2188, 2013.

\bibitem{rollings_natcomms_2016}
R.~C. Rollings, A.~T. Kuan, and J.~A. Golovchenko.
\newblock Ion selectivity of graphene nanopores.
\newblock {\em Nature Comm.}, 7(1):11408, 2016.

\bibitem{rangharajan_micronano_2016}
K.~K. Rangharajan, M.~Fuest, A.~T. Conlisk, and S.~Prakash.
\newblock Transport of multicomponent, multivalent electrolyte solutions across
  nanocapillaries.
\newblock {\em Microfluid. Nanofluid.}, 20(4):54, 2016.

\bibitem{nandigana_scirep_2018}
V.~V.~R. Nandigana, K.~Jo, A.~Timperman, and N.~R. Aluru.
\newblock Asymmetric-fluidic-reservoirs induced high rectification nanofluidic
  diode.
\newblock {\em Sci. Rep.}, 8(1):13941, 2018.

\bibitem{wang_jphyschemc_2018}
X.~Wang, Y.~Chen, Z.~Meng, Q.~Zhang, and J.~Zhai.
\newblock Effect of trivalent
  {\textquotedblleft}calcium-like{\textquotedblright} cations on ionic
  transport behaviors of artificial calcium-responsive nanochannels.
\newblock {\em J. Phys. Chem. C}, 122(43):24863--24870, 2018.

\bibitem{nasir_jcis_2019}
S.~Nasir, M.~Ali, J.~Cervera, V.~Gomez, M.~H.~A. Haider, W.~Ensinger, S.~Mafe,
  and P.~Ramirez.
\newblock Ionic transport characteristics of negatively and positively charged
  conical nanopores in 1:1, 2:1, 3:1, 2:2, 1:2, and 1:3 electrolytes.
\newblock {\em J. Coll. Interf. Sci.}, 553:639--646, 2019.

\bibitem{alidoosti_electrophoresis_2019}
E.~Alidoosti and H.~Zhao.
\newblock The effects of electrostatic correlations on the ionic current
  rectification in conical nanopores.
\newblock {\em Electrophoresis}, 2019.

\bibitem{liu_nnl_2019}
C.~Liu, T.~Ding, L.~Wu, Z.~Meng, and Z.~Lu.
\newblock Adsorption of monovalent cations and surface charge inversion within
  nanopores studied by ion current response and {Poisson-Nernst-Planck}
  simulations.
\newblock {\em Nanosci. Nanotech. Lett.}, 11(2):205--213, 2019.

\bibitem{li_jphyschemc_2019}
Z.~Li, Y.~Qiu, Y.~Zhang, M.~Yue, and Y.~Chen.
\newblock Effects of surface trapping and contact ion pairing on ion transport
  in nanopores.
\newblock {\em J. Phys. Chem. C}, 123(24):15314--15322, 2019.

\bibitem{zqli_jphyschemc_2019}
Z-Q. Li, Y.~Wang, Z-Q. Wu, M-Y. Wu, and X-H. Xia.
\newblock Bioinspired multivalent ion responsive nanopore with ultrahigh ion
  current rectification.
\newblock {\em J. Phys. Chem. C}, 123(22):13687--13692, 2019.

\bibitem{zhang_analchem_2019}
X.~Zhang, X.~Han, S.~Qian, Y.~Yang, and N.~Hu.
\newblock Tuning ion transport through a nanopore by self-oscillating chemical
  reactions.
\newblock {\em Anal. Chem.}, 91(7):4600--4607, 2019.

\bibitem{boda_jctc_2012}
D.~Boda and D.~Gillespie.
\newblock Steady state electrodiffusion from the {Nernst-Planck} equation
  coupled to {Local Equilibrium Monte Carlo} simulations.
\newblock {\em J. Chem. Theor. Comput.}, 8(3):824--829, 2012.

\bibitem{boda_jml_2014}
D.~Boda, R.~Kov\'acs, D.~Gillespie, and T.~Krist\'of.
\newblock Selective transport through a model calcium channel studied by
  {Local} {Equilibrium} {Monte} {Carlo} simulations coupled to the
  {Nernst}-{Planck} equation.
\newblock {\em J. Mol. Liq.}, 189:100--112, 2014.

\bibitem{matejczyk_jcp_2017}
B.~Matejczyk, M.~Valisk{\'{o}}, M.-T. Wolfram, J.-F. Pietschmann, and D.~Boda.
\newblock Multiscale modeling of a rectifying bipolar nanopore: {Comparing}
  {Poisson-Nernst-Planck} to {Monte Carlo}.
\newblock {\em J. Chem. Phys.}, 146(12):124125, 2017.

\bibitem{gillespie-jpcb-109-15598-2005}
D.~Gillespie, L.~Xu, Y.~Wang, and G.~Meissner.
\newblock {({De})constructing the Ryanodine Receptor: {Modeling} Ion Permeation
  and Selectivity of the Calcium Release Channel}.
\newblock {\em J. Phys. Chem. B}, 109(32):15598--15610, 2005.

\bibitem{gillespie_bj_2008_energetics}
D.~Gillespie.
\newblock Energetics of divalent selectivity in a calcium channel: {The}
  {Ryanodine Receptor} case study.
\newblock {\em Biophys. J.}, 94(4):1169--1184, 2008.

\bibitem{gillespie_bj_2008}
D.~Gillespie and D.~Boda.
\newblock The anomalous mole fraction effect in calcium channels: {A} measure
  of preferential selectivity.
\newblock {\em Biophys. J.}, 95(6):2658--2672, 2008.

\bibitem{gillespie_bj_2008b}
D.~Gillespie, D.~Boda, Y.~He, P.~Apel, and Z.S. Siwy.
\newblock Synthetic nanopores as a test case for ion channel theories: {The}
  anomalous mole fraction effect without single filing.
\newblock {\em Biophys. J.}, 95(2):609--619, 2008.

\bibitem{boda_jgp_2009}
D.~Boda, M.~Valisk{\'o}, D.~Henderson, B.~Eisenberg, D.~Gillespie, and
  W.~Nonner.
\newblock Ionic selectivity in {L-Type} calcium channels by electrostatics and
  hard-core repulsion.
\newblock {\em J. Gen. Physiol.}, 133(5):497--509, 2009.

\bibitem{boda_arcc_2014}
D.~Boda.
\newblock In R.~A. Wheeler, editor, {\em Ann. Rep. Comp. Chem.}, volume~10,
  chapter 5 {Monte Carlo} Simulation of Electrolyte Solutions in Biology: {In}
  and Out of Equilibrium, pages 127--163. Elsevier, 2014.

\bibitem{madai_jcp_2017}
E.~M\'adai, M.~Valisk\'o, A.~Dallos, and D.~Boda.
\newblock Simulation of a model nanopore sensor: {Ion} competition underlines
  device behavior.
\newblock {\em J. Chem. Phys.}, 147(24):244702, 2017.

\bibitem{blum_mp_1975}
L.~Blum.
\newblock Mean spherical model for asymmetric electrolytes.
\newblock {\em Mol. Phys.}, 30(5):1529--1535, 1975.

\bibitem{blum_jcp_1977}
L.~Blum and J.~S. Hoeye.
\newblock Mean spherical model for asymmetric electrolytes. 2. {Thermodynamic}
  properties and the pair correlation function.
\newblock {\em J. Phys. Chem.}, 81(13):1311--1316, 1977.

\bibitem{2000_nonner_bj_1976}
W.~Nonner, L.~Catacuzzeno, and B.~Eisenberg.
\newblock {Binding and Selectivity in {L}-Type Calcium Channels: {A} Mean
  Spherical Approximation.}
\newblock {\em Biophys. J.}, 79(4):1976--1992, 2000.

\bibitem{valisko_jml_2007}
M.~Valisk{\'o}, D.~Henderson, and D.~Boda.
\newblock The capacitance of the electrical double layer of valence-asymmetric
  salts at low reduced temperatures.
\newblock {\em J. Mol. Liq.}, 131--132:179--184, 2007.

\bibitem{west2017scale}
G.B. West.
\newblock {\em Scale}.
\newblock Orion Publishing Group, Limited, 2017.

\end{thebibliography}

\end{document}